\documentclass[]{aastex701}

\usepackage{graphicx}
\usepackage{subcaption}
\usepackage{gensymb}

\begin{document}

\title{Complete evolution of a blue straggler (WOCS 5011) and white dwarf system of NGC 7789 using MESA}

\author{Anju Panthi}
\affiliation{Indian Institute of Astrophysics, Sarjapur Road, Koramangala, Bangalore 560034, India}
\email{p20190413@pilani.bits-pilani.ac.in}

\author{Annapurni Subramaniam}
\affiliation{Indian Institute of Astrophysics, Sarjapur Road, Koramangala, Bangalore 560034, India}
\email{purni@iiap.res.in}

\author{S. R. Dhanush}
\affiliation{Indian Institute of Astrophysics, Sarjapur Road, Koramangala, Bangalore 560034, India}
\email{}

\author{Robert D. Mathieu}
\affiliation{Department of Astronomy, University of Wisconsin-Madison, 475 N Charter St, Madison, WI 53706, USA}
\email{}

\author{Kaushar Vaidya}
\affiliation{Department of Physics, Birla Institute of Technology and Science - Pilani, Pilani 333031, Rajasthan, India}
\email{}

\begin{abstract}
We present a binary evolution model of the blue straggler star (BSS; WOCS 5011) in open cluster NGC 7789 using the Modules for Experiments in Stellar Astrophysics (MESA). Starting from a primordial binary consisting of a $\sim 1.85 M_{\odot}$ donor (white dwarf progenitor) and a $\sim 1.67 M_{\odot}$ accretor (BSS progenitor) with an initial orbital period of 463 days, the model yields a final period of $\sim 2708$ days, consistent with the observed value of $2710 \pm 60$ days. The donor evolves through the red giant and asymptotic giant branch phases, transferring mass through wind mass transfer, wind Roche-lobe overflow (WRLOF), and a brief episode of Roche-lobe overflow (RLOF). The accretor gains $\sim 0.50 M_{\odot}$ to become a $2.17 M_{\odot}$ BSS, while the donor loses its envelope to form a $0.55,M_{\odot}$ carbon--oxygen white dwarf. The overall mass-transfer efficiency is $\sim 38\%$. Although WRLOF operates over most of the evolution, the short-lived TP--AGB RLOF phase transfers the largest amount of mass, with an efficiency of $\sim$ 35$\%$. The model reproduces the observed BSS properties and predicts a white dwarf with $T_{\rm eff}=12{,}520$ K, $R=0.013 R_{\odot}$, and $L=0.004 L_{\odot}$. A revised two-component SED fit yields a hot companion with $T_{\rm eff}=13{,}250 \pm 250$ K, $R=0.32 \pm 0.07 R_{\odot}$, and $L=2.86^{+2.19}_{-1.81} L_{\odot}$. While the white dwarf temperature is comparable to the MESA prediction, the inferred radius and luminosity are larger, supporting a recently stripped system.
\end{abstract}

\keywords{\uat{Blue straggler stars}{168} --- \uat{Binary stars}{154} --- \uat{Close binary stars}{254} --- \uat{Open star clusters}{1160} --- \uat{Stellar evolution}{1599} --- \uat{White dwarf stars}{1799}}

\section{Introduction} 

Most stars form and evolve in binary or higher-order multiple systems \citep{moe2017mind,raghavan2010survey}. Binary evolution, through mass transfer, mergers, or dynamical perturbations, fundamentally differs from single-star stellar evolution pathways \citep{eggleton2006evolutionary}. Blue straggler stars (BSSs), first identified by \citet{sandage1953color}, are among the products of such alternative evolutionary pathways. Historically, BSSs have been identified photometrically as stars lying brighter and bluer than the main-sequence turn-off (MSTO) of a coeval stellar population. More broadly, however, BSSs are main-sequence stars that have gained mass during their evolution and continue to undergo core hydrogen burning; consequently, a mass-gaining star need not necessarily lie brighter than the MSTO \citep{mathieu2025blue}. Such stars are generally more massive than the normal turn-off population, and their enhanced masses can also lead to a more centrally concentrated spatial distribution in dynamically evolved stellar systems \citep{lauzeral1992radial}.
Several formation pathways have been proposed to explain the origin of BSSs, with three mechanisms most commonly discussed in the literature. One possibility involves direct stellar collisions between two or more stars, which are expected to occur primarily in dense stellar environments such as cores of the globular clusters (GCs; \citealt{chatterjee2013stellar}). Another widely accepted scenario is mass transfer within binary systems, where material is transferred from an evolved donor star to its companion through Roche-lobe overflow (RLOF; \citealt{mccrea1964extended}), classical wind accretion (e.g., Bondi–Hoyle–Lyttleton; \citealt{bondi1944mechanism}), or efficient wind Roche-lobe overflow (WRLOF; \citealt{mohamed2007astronomical}), typically producing a BSS in a binary system and less often a single BSS. The binary mass transfer mechanism is commonly divided into three cases, A, B, and C, depending on the evolutionary stage of the donor star at the onset of mass transfer. In Case A, mass transfer occurs while the donor is still on the main sequence, producing either a merged BSS or a close binary containing a BSS and a main-sequence companion with a short orbital period of $\sim0.5$–10 days \citep{webbink1976evolution}. In Case B, the mass transfer begins after the donor has evolved off the main sequence and ascended the red giant branch (RGB). This channel generally produces binaries with orbital periods of roughly $\sim10$–300 days and BSS with helium WD companions of masses $\lesssim0.45\,M_{\odot}$ \citep{chen2008mass}. Finally, in Case C, mass transfer occurs when the donor star is on the asymptotic giant branch (AGB), typically leading to BSS in long-period binaries with orbital periods of $\sim300$–3000 days that host carbon–oxygen white dwarf (WD) companions of masses around $\sim0.5\,M_{\odot}$ \citep{chen2008mass,gosnell2019constraining}. A third pathway involves the evolution of hierarchical triple systems, where dynamical interactions can induce mass transfer or mergers in the inner binary, leading to the formation of BSSs \citep{perets2009triple}.

Ultraviolet (UV) observations using HST \citep{ford1998advanced}, UVOT \citep{roming2005swift}, and UVIT \citep{agrawal2006broad} have provided direct evidence for hot companions to BSSs in open clusters (OCs). For example, far-UV observations of NGC~188 using HST revealed WD companions to several BSSs, conclusively establishing mass transfer as a dominant formation channel \citep{gosnell2014detection, gosnell2015implications, gosnell2019constraining}. Within the UVIT Open Cluster Survey (UOCS), the first clear result came from NGC~188, where UV excess in the multiwavelength spectral energy distribution (SED) of a BSS revealed a hot post-AGB/HB companion, demonstrating recent mass transfer in a BSS binary \citep{subramaniam2016hot}. Successive UVIT studies identified multiple low-mass (LM) and extremely low-mass (ELM) WD companions to BSSs in M67, establishing that a significant fraction of the BSS population consists of post-mass-transfer systems \citep{jadhav2019uvit, sindhu2019uvit, pandey2021uocs}. In King~2, six UV-bright BSSs were found to host hot companions with properties similar to EHB/sdB stars \citep{jadhav2021blue}; in NGC~7789, five out of 15 BSS candidates were found to host ELM WDs, confirming post-mass-transfer formation in at least one-third of the sample \citep{vaidya2022uocs}; in NGC~2506, hot companions were identified in three BSSs, while the broader exotic population revealed ELM, LM, normal-mass, and high-mass WD companions, indicating that both mass-transfer and merger pathways are active \citep{panthi2022uocs}; in NGC~7142, several BSSs show UV excess with hot companions ranging from ELM to high-mass WDs, providing strong evidence that multiple formation channels contribute \citep{panthi2024uocs}; in NGC~2627, binary-component SED fits indicate ELM WD companions to BSSs and other exotic systems \citep{saketh2024uocs}; and in NGC~2420, UVIT identified FUV-bright BSS systems, including a BSS+ELM WD binary, consistent with a recent post-mass-transfer origin \citep{yadav2024uocs}. Complementing these wide-field UVIT studies, Swift/UVOT observations have also been successfully used to characterize BSSs in OCs through multiwavelength SED analysis. \citet{rao2022characterization} reported a low-mass WD companion to a BSS in Melotte~66, while \citet{chand2024characterization} and \citet{sheikh2024characterization} identified ELM WD companions to BSSs in Berkeley~39 and NGC~2243, respectively.
 
The BSS WOCS 5011 (RA = 359.24595 deg , DEC = +56.79255 deg) in NGC~7789 represents a particularly compelling system for investigating BSS formation through binary evolution. It is a long-period spectroscopic binary with P  = 2710 $\pm$ 60 days and e = 0.32 $\pm$ 0.04 \citep{nine2020wiyn}, exhibits significant Barium (Ba) enhancement indicative of past mass transfer from an AGB companion, and is identified as a rapid rotator \citep{nine2024wiyn}. Such a combination of orbital, chemical, and rotational properties provides strong evidence for a mass-transfer origin. However, despite being identified as a single-lined spectroscopic binary (SB1) system and a confirmed binary member \citep{nine2020wiyn, nine2024wiyn}, UVIT observations did not reveal any UV excess for this object \citep{vaidya2022uocs}, suggesting the absence of a detectable hot companion. The coexistence of a wide orbit, chemical enrichment, and the lack of a UV-detected companion therefore poses important constraints on the nature, timing, and efficiency of mass transfer, hence an important system for detailed binary evolution analysis. In this work, we follow the analysis pathway of \cite{sun2023wocs} for the BSS WOCS 4540. Detailed analysis are necessary to throw light on the details of mass transfer in various evolutionary phases of the binary system in order to connect to the outcome of such evolution.

The structure of this paper is as follows. In Section~\ref{Section2}, we describe the observed properties of the OC NGC~7789 and the binary system WOCS~5011, which serve as the key constraints for our modeling. In Section~\ref{Section3}, we outline the binary evolution modeling approach adopted in this work. In Section~\ref{Section4}, we present the detailed evolutionary history of the system. In Section~\ref{Section5}, we discuss the physical implications of our results and compare our findings with previous studies. Finally, in Section~\ref{Section6}, we summarize our main results.

\section{NGC 7789 and WOCS 5011 observed properties} \label{Section2}

NGC 7789 ($\alpha_{\mathrm{J2000}} = 23^{\mathrm h}57^{\mathrm m}19^{\mathrm s}$, $\delta_{\mathrm{J2000}} = +56^{\circ}43^{\prime}28^{\prime\prime}$) is a rich intermediate-age OC located in the constellation Cassiopeia that has been extensively studied in the literature \citep{friel2002metallicities, nagarajan2023chemical, anthony2026wocs}. The cluster lies toward the Galactic anticentre at Galactic coordinates $(l,b)\sim(115.5^{\circ},-5.4^{\circ} $ ; \citealt{ferreira2025open}). Owing to its rich population and favourable location in the Galactic disk, NGC 7789 serves as an important intermediate-age benchmark system for investigations of stellar evolution and the chemical evolution of the Galactic disk. The cluster has a clearly defined main-sequence turnoff, an extended RGB, and a remarkably large population of BSSs \citep{friel2002metallicities, anthony2026wocs}. Based on deep \textit{Gaia} photometry, the cluster is estimated to host up to $\sim$ 3000 probable members down to $G\sim21$ mag \citep{gao2018memberships}, making it one of the richest known intermediate-age OCs in the solar neighbourhood. Early photometric investigations of NGC 7789 date back to the work of \citet{burbidge1958color}, who obtained $UBV$ photometry for nearly 700 stars and derived an initial reddening estimate of $E(B-V)=0.24$. Subsequent studies incorporating modern photometric datasets and \textit{Gaia}-based membership analyses have revised the reddening to higher values. For example, \citet{anthony2026wocs} derived $E(B-V)=0.30\pm0.02$, while \citet{ferreira2025open} obtained $E(B-V)=0.328\pm0.008$ from \textit{Gaia} DR3 data. The cluster’s relatively low Galactic latitude results in substantial interstellar extinction, which has historically complicated the interpretation of its photometric properties. Earlier photometric analyses suggested somewhat subsolar metallicity values, such as \big[Fe/H\big] $=-0.26\pm0.10$ \citep{friel2002metallicities}. However, high-resolution spectroscopic studies now consistently indicate near-solar composition, including \big[Fe/H\big] $=-0.04\pm0.05$ \citep{tautvaivsiene2005chemical} and \big[Fe/H\big] $=+0.02\pm0.04$ \citep{jacobson2011chemical}. Detailed abundance analyses also reveal significant enhancements in neutron-capture elements, particularly Ba, with reported values of \big[Ba/Fe\big] $\sim 0.47 $–$ 0.48$ \citep{pancino2010chemical, overbeek2015erratum}. In addition, mild $\alpha$-element enhancement and characteristic C–N–O abundance patterns have been observed, reflecting internal mixing processes and dredge-up effects during stellar evolution \citep{overbeek2015erratum,nagarajan2023chemical}. Age estimates for NGC 7789 show some dispersion in the literature. Classical photometric studies based on CMD fitting typically report ages between 1.4 and 1.6 Gyr \citep{gim1998open,ferreira2025open}. Using homogeneous isochrone fitting with \textit{Gaia} DR2 astrometry, \citet{cantat2020painting} derived a somewhat younger age of $\sim1.2$ Gyr. The apparent spread among published values largely reflects differences in methodology, as well as uncertainties associated with extinction corrections. Observational evidence for an extended main-sequence turnoff (eMSTO) has also been reported. For example, \citet{de2020extended} interpreted the width of the MSTO as corresponding to an age dispersion of roughly 390 Myr, yielding a representative age of $1070\pm390$ Myr. However, the physical interpretation of the eMSTO remains debated and may instead be explained by stellar rotation, convective overshooting, or differential reddening \citep{anthony2026wocs}. Overall, the range of age estimates is thought to arise primarily from a combination of differential reddening, stellar rotation, and a significant binary fraction within the cluster. \cite{nine2020wiyn} identified 624 probable cluster members and determined orbital solutions for 81 spectroscopic binaries with orbital periods spanning $\sim$ 1.45 to 4200 days under WOCS. Recently, \citet{yakut2026binary} determined the fundamental parameters of NGC 7789 by analysing six well-characterized binary systems using RV measurements, \textit{TESS} photometry, and broadband SEDs. By combining dynamical constraints with the observed SEDs, their joint modelling yielded a cluster age of $1.26\pm0.09$ Gyr. This value lies toward the younger end of, but remains broadly consistent with, the commonly adopted age range of $\sim1.2$–$1.6$ Gyr obtained from recent \textit{Gaia}-based isochrone analyses \citep{dias2021updated,ferreira2025open}.

The BSS WOCS 5011 in NGC 7789 exhibits a significant Ba enhancement with \big[Ba/Fe\big] = 1.32 \citep{nine2024wiyn}. The enhancement of slow neutron-capture (s-process) elements, such as Ba, can arise from nucleosynthesis during the thermally pulsing (TP) phase of AGB evolution \citep{iben1983asymptotic, busso1999nucleosynthesis}. Such Ba enrichments have been widely observed in field binary systems, both among giant stars \citep{jorissen2019barium} and dwarf stars \citep{escorza2019barium}. Furthermore, evidence of past mass transfer leading to extrinsic s-process enrichment has also been identified in field binary systems containing AGB stars \citep{shetye2020discovery, shetye2021s}. \citet{nine2020wiyn} reported this system to be a SB1 with period a 2710$\pm$60 days and eccentricity 0.32$\pm$0.04. 
\cite{nine2024wiyn} examine Ba abundances in BSSs in two OCs, including NGC 7789, to test for chemical signatures of mass transfer from AGB companions. They suggested WOCS 5011 to be Ba-enriched and also reported it a rapid rotator (vsin$\textit{i}$ 50 km/s). In the standard Ba-star framework described by \cite{escorza2023barium}, this enrichment is naturally explained if the companion previously evolved through the TP-AGB phase, synthesized s-process elements, and transferred enriched material before contracting to a WD remnant. \cite{nine2024wiyn} also estimate the effective temperatures of their sample by first producing continuum-normalized H$\alpha$ spectra using the Python package \texttt{specutils} \citep{earl2022astropy}. Then they compared the spectra to a grid of theoretical, solar-metallicity H$\alpha$ profiles from \citep{barklem2002detailed}, spanning surface gravities of 3.6 $\leq$ log${10}$(g) $\leq$ 4.5 and effective temperatures in the range 5000–-7500 K. Surface gravities were estimated using the \texttt{starfit} function from the \texttt{isochrones} package \citep{mortON2015isochrones}, which interpolates across MIST isochrones \citep{dotter2016mesa, choi2016mesa}. For the star WOCS 5011, they report T$_{\mathrm{eff}}$ = 8350$^{+140}_{-130}$ K and $\log g$ = 3.66$^{+0.07}_{-0.06}$. UV photometric observations of NGC 7789 have been carried out using the UVIT onboard \textit{AstroSat} \citep{agrawal2006broad, kumar2012ultraviolet}.  \cite{vaidya2022uocs} reported WOCS 5011 as a single star with no UV excess having $\log g$ 4.0, T$_{\mathrm{eff}}$ 8750 $\pm$ 125 K, radius 3.04 $\pm$ 0.152 R$_{\odot}$, and luminosity 47.7 $\pm$ 4.81 L$_{\odot}$.

\section{Finding the best-fit model for WOCS 5011} \label{Section3}

\subsection{Methodology}

We performed detailed binary evolution modeling of WOCS 5011 using the Modules for Experiments in Stellar Astrophysics code
(MESA; \citealt{paxton2011modules,paxton2013modules, paxton2015modules, paxton2017modules, paxton2019modules}). 
The binary evolution calculations presented in this work assume a circular orbit ($e=0$). This differs from the observed present-day eccentricity of WOCS~5011, $e=0.32\pm0.04$. Thus, the orbital evolution and mass-transfer history derived here should be interpreted within the limitations of the circular-orbit approximation. In an eccentric binary, the instantaneous
binary separation, and consequently the Roche-lobe geometry, vary over the orbital cycle, which may affect the onset and efficiency of mass transfer.

 We adopt an approach followed for the NGC 188 BSS binary WOCS 4540 by \citet{sun2023wocs}. In this framework, the initially more massive primary star evolves to form the present-day WD, while the initially less massive secondary accretes mass and becomes the BSS. We assume that the two stars have formed together in a primordial binary and evolve simultaneously. The simulations were carried out using the \texttt{evolve\_both\_stars} binary module in MESA. In addition to the standard Roche-lobe overflow (RLOF) treatment implemented in MESA, we include the wind Roche-lobe overflow (WRLOF), which is expected to operate when the slow, dense winds from an AGB donor are gravitationally focused toward the companion \citep{mohamed2010understanding,abate2013wind}. Wind mass transfer is enabled throughout the binary evolution using a custom wind-transfer routine based on the WRLOF prescription, replacing the default Bondi--Hoyle implementation in MESA. During the RGB phase, the donor loses mass through stellar winds that are only weakly accreted by the companion. As the donor evolves onto the AGB and TP-AGB, the WRLOF mechanism becomes increasingly efficient, substantially enhancing the wind-accretion rate. Once the donor fills its Roche lobe, mass transfer proceeds through the standard RLOF formalism implemented in MESA.

We allow non-conservative mass transfer, such that a fraction of the mass lost by the donor escapes from the system, and define the overall mass-transfer efficiency as the ratio of the total mass gained by the accretor and the total mass lost by the donor. The mass lost from the system is treated using the $\beta$-mechanism, in which $\beta$ denotes the fraction of the transferred mass that escapes from the vicinity of the accretor. The stellar winds from the donor are implemented using the Reimers prescription \citep{baschek2012problems} during the RGB phase and the Blocker scheme \citep{blocker1995stellar} during the AGB phase. We adopt Z to be 0.015 \citep{jacobson2011chemical, overbeek2015ngc}. During the evolution, we keep the accretor within its Roche lobe. If the accretor overfills its Roche lobe, the calculation is terminated because the subsequent contact evolution, in which both stars fill their Roche lobes, is not followed in the present modeling. We note that contact does not necessarily imply the immediate onset of common-envelope evolution \citep{kolb1990comparative}. In the following description, quantities associated with the donor star (the progenitor of the WD) are denoted with the subscript “d,” while those corresponding to the accretor star (the progenitor of the BSS) are denoted with the subscript “a.”

Mass transfer plays a crucial role in shaping the evolutionary pathways of stars in binary systems \citep{sana2012binary,de2017dawes}. Material can be transferred from the donor to the accretor through two main channels. In close binaries, when the donor expands sufficiently to fill its Roche-lobe radius ($R_{\mathrm{L},d}$), matter can flow directly toward the companion through RLOF. In wider systems, some of the material lost through stellar winds from the donor may instead be captured by the companion, resulting in wind-driven mass transfer \citep{bondi1944mechanism}. An intermediate mode of mass transfer, known as WRLOF, occurs when the stellar wind of the donor is confined within the wind acceleration region surrounding the star. If the wind acceleration radius becomes comparable to or larger than $R_{\mathrm{L},d}$, the slow outflow can be gravitationally focused toward the companion through the inner Lagrangian point where the Roche lobes of the two stars meet \citep{mohamed2007astronomical}. This mechanism is particularly relevant for systems containing AGB stars, which are characterized by slow and dense winds \citep{abate2013wind}. In these stars, pulsations in the extended atmosphere generate shocks that lift material to large radii, where cooling allows dust grains to condense. Radiation pressure acting on these grains then accelerates the gas–dust mixture outward, producing slow winds with typical mass-loss rates of $\sim10^{-7}$–$10^{-5}\,M_{\odot}\,\mathrm{yr^{-1}}$ and wind velocities of $\sim10$–$20\,\mathrm{km\,s^{-1}}$ \citep{herwig2005evolution,habing2013asymptotic}. 

We implement WRLOF following the \citet{abate2013wind} prescription, similar to the approach adopted by \citet{sun2023wocs}. In this formalism, the accretion efficiency of the stellar wind depends on the ratio between the wind acceleration radius and the Roche-lobe radius of the donor star. The WRLOF accretion efficiency is expressed as

\begin{equation}
\beta_{\rm acc} = c_1 \left(\frac{R_{\rm wind}}{R_{\rm L,d}}\right)
+ c_2 \left(\frac{R_{\rm wind}}{R_{\rm L,d}}\right)^2
+ c_3 ,
\end{equation}

where $R_{\rm L,d}$ is the Roche-lobe radius of the donor star and $R_{\rm wind}$ denotes the radius of the wind acceleration zone. We adopt the values of fitting coefficients, $c_1=-0.284$, $c_2=0.918$, and $c_3=-0.234$ following \cite{abate2013wind}.

The wind acceleration radius is estimated using

\begin{equation}
R_{\rm wind} = R_d
\left(\frac{T_{\rm eff}}{T_{\rm cond}}\right)^{2.5},
\end{equation}

where $R_d$ is the radius of the donor star, $T_{\rm eff}$ is its effective temperature, and $T_{\rm cond}$ is the condensation temperature of dust grains. We adopt $T_{\rm cond}=1500\,{\rm K}$, appropriate for carbon-rich dust.

We further include the mass-ratio dependence described by \citet{abate2013wind} when computing the WRLOF accretion efficiency. The final efficiency is therefore given by

\begin{equation}
\beta_{\rm acc} = \min \left( \frac{25}{9} q^2 \beta_{\rm acc}, \,
\beta_{\rm acc,max} \right),
\end{equation}

where $q$ is the binary mass ratio and $\beta_{\rm acc,max}$ is the maximum allowed WRLOF efficiency. We adopt $\beta_{\rm acc,max}=0.5$, consistent with the upper limit obtained from hydrodynamical simulations \citep{mohamed2010understanding}. Since the WRLOF prescription of \citet{abate2013wind} was developed for the AGB phase, we apply this formalism only after the central helium mass fraction drops below $10^{-4}$, corresponding to the end of core helium burning. For wind mass transfer during the RGB phase, we assume a constant mass-transfer efficiency of 50\%, corresponding to the maximum WRLOF efficiency inferred from simulations by \citet{mohamed2010understanding}. 

\subsection{Model Implementation and Best-fit Binary Model}

The best-fitting binary model for WOCS 5011 was obtained through a multi-stage grid search of initial binary parameters using \texttt{MESA} binary evolution calculations. The initial parameter ranges were motivated by the observed properties of the cluster NGC 7789, particularly its age of $\sim$1.8~Gyr and a main-sequence turnoff mass of $\sim$1.55~$M_{\odot}$. In the first grid, we vary the donor mass ($M_d$) between $1.8$--$2.0~M_{\odot}$ (step size $0.05~M_{\odot}$), the accretor mass ($M_a$) between $1.4$--$1.75~M_{\odot}$ (step size $0.05~M_{\odot}$), the initial orbital period between $10$--$1000$ days (step size $30$ days), and the RLOF mass-transfer efficiency parameter $\beta$ between $0.3$--$0.95$ (step size $0.05$). From this initial search, only a small subset of models produced final orbital periods close to the observed value of $2710 \pm 60$ days. We then performed a second round of calculations to refine the parameter space around the solutions obtained from the first grid. The parameter space was re-explored with finer steps of $0.02~M_{\odot}$ in $M_d$, $0.01~M_{\odot}$ in $M_a$, an initial orbital period between $410$--$510$ days (step size $4$ days), and $0.02$ in $\beta$. Then a third grid was computed to quantify the uncertainties in the optimized parameters and determine the best-fit model. This final tuning employed step sizes of $0.01~M_{\odot}$ for both stellar masses, an initial orbital period between $457$--$463$ days (step size $1$ day), and $0.01$ for the RLOF mass-transfer efficiency. Finally, we determined the best-fit model by performing a $\chi^{2}$ test and selected the model that yields the minimum $\chi^{2}$ value between the observed and model parameters. The best-fit model parameters, including the initial donor mass $M_{d,i}$, initial accretor mass $M_{a,i}$, initial orbital period $P_{\mathrm{orb},i}$, the RGB mass-transfer efficiency $\beta$, Reimers scaling factor, and the Blöcker scaling factor, are listed in Table \ref{Table1}.

\begin{table}[t]
\centering
\caption{Initial parameters of the best-fit binary model.}
\begin{tabular}{lc}
\hline
Parameter & Value \\
\hline
Initial donor mass, $M_{d,i}$ ($M_\odot$) & 1.85 \\
Initial accretor mass, $M_{a,i}$ ($M_\odot$) & 1.67 \\
Initial orbital period, $P_{\rm orb,i}$ (days) & 463 \\
RLOF efficiency, $\beta$ & 0.35 \\
Reimers scaling factor, $\eta_{\rm R}$ & 0.5 \\
Bl\"ocker scaling factor, $\eta_{\rm B}$ & 0.1 \\
\hline
\end{tabular}
\label{Table1}
\end{table}

\section{The best-fit model} \label{Section4}
\subsection{The evolution of WOCS 5011 on the Color-Magnitude Diagram}

\begin{figure*}[h]
\centering
\begin{subfigure}{0.48\textwidth}
\centering
\includegraphics[width=\linewidth]{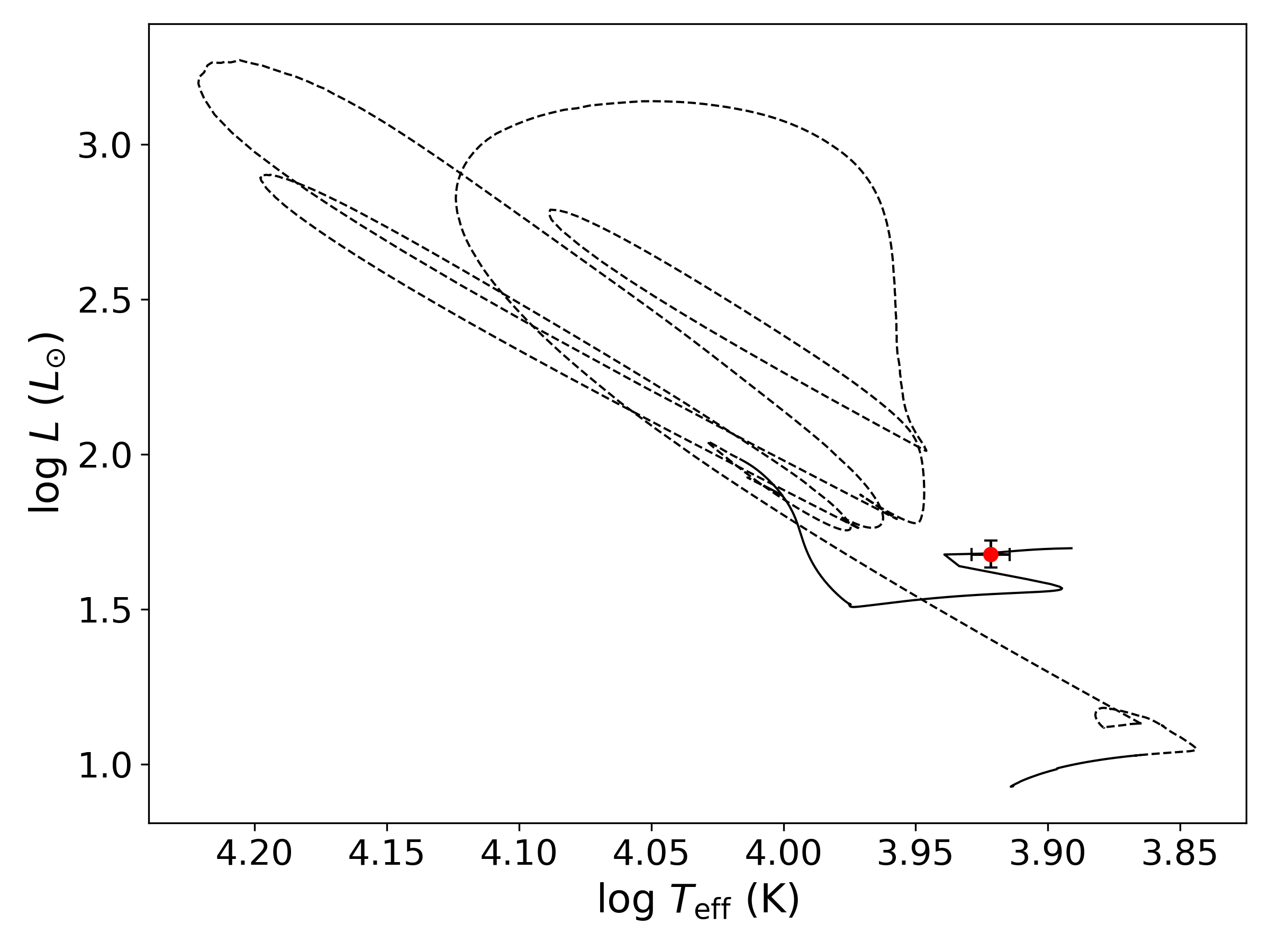}
\caption{}
\label{Figure1a}
\end{subfigure}
\hfill
\begin{subfigure}{0.48\textwidth}
\centering
\includegraphics[width=\linewidth]{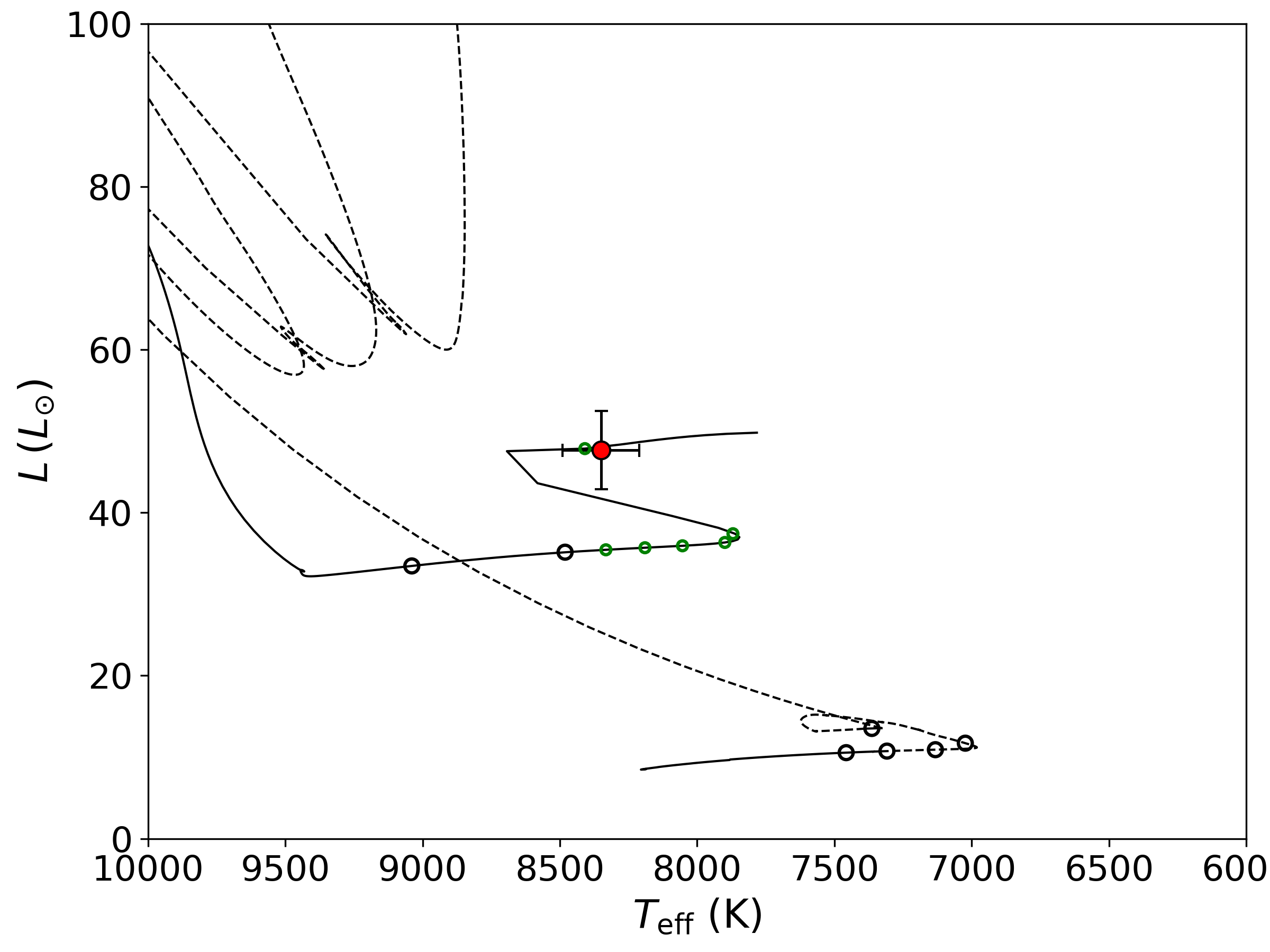}
\caption{}
\label{Figure1b}
\end{subfigure}

\caption{(a) Evolutionary track of the accretor (BSS) in the $\log T_{\mathrm{eff}}$--$\log L$ plane. Solid and dashed segments denote phases of negligible ($\dot{M} < 10^{-12}\,M_{\odot}\,\mathrm{yr}^{-1}$) and active ($\dot{M} > 10^{-12}\,M_{\odot}\,\mathrm{yr}^{-1}$) mass transfer, respectively. The red point with error bars shows the observed position of WOCS 5011. Loop-like features during accretion reflect structural readjustments, while the post-accretion evolution is smooth and occurs at higher luminosity due to mass gain. (b) Zoomed-in view near the observed position. Black open circles indicate model snapshots between $1.0$--$1.6\,\mathrm{Gyr}$, with a time spacing of $0.1\,\mathrm{Gyr}$, and green open circles mark finer sampling between $1.62$--$1.72\,\mathrm{Gyr}$, with a time spacing of $0.02\,\mathrm{Gyr}$. The observed position reflects a post-accretion, detached phase at $\sim 1.72\,\mathrm{Gyr}$.}
\label{Figure1}
\end{figure*}

Figure \ref{Figure1a} presents the evolutionary trajectory of the accretor (BSS progenitor) in the $\log T_{\mathrm{eff}}$--$\log L$ plane. Solid segments correspond to phases in which the mass-transfer rate is negligible ($\dot{M} < 10^{-12}\,M_{\odot}\,\mathrm{yr}^{-1}$), while dashed segments denote epochs of active accretion ($\dot{M} > 10^{-12}\,M_{\odot}\,\mathrm{yr}^{-1}$). The red point with error bars marks the observed position of WOCS 5011 with reported uncertainities in temperature and luminosity \citep{vaidya2022uocs, nine2024wiyn}.

According to the model, the accretor increases mass from $1.67\,M_{\odot}$ to $2.17\,M_{\odot}$. %The elevated luminosity in the post-accretion phase is therefore consistent with the increased stellar mass, as expected from the mass--luminosity relation for hydrogen-burning stars in this mass range.
Figure \ref{Figure1b} presents a zoomed-in view of Figure \ref{Figure1a}, concentrating on the region of the T$_{eff}$ - L plane surrounding the observed position of the accretor. Black open circles indicate model snapshots between $1.0$--$1.6\,\mathrm{Gyr}$ (spaced by $0.1\,\mathrm{Gyr}$), while green open circles mark finer age markers between $1.62$--$1.72\,\mathrm{Gyr}$ (with $0.02\,\mathrm{Gyr}$ spacing). The observed BSS position lies on this relaxed, post-accretion portion of the track and coincides with model ages near 1.72 Gyr, as indicated by the green open circles. The quantitative agreement between model predictions and observations of the BSS properties is summarized in Table \ref{Table2}. The model suggests a WD companion of mass $0.55\,M_{\odot}$, of $T_{\mathrm{eff}}\,12{,}520\,\mathrm{K}$, and of $\log g\,7.92$.

\begin{table}[h]
\centering
\caption{Comparison between observed properties and best-fit model of the BSS, along with the binary orbital period.}
\begin{tabular}{lcc}
\hline
Parameter & Observed & Model \\
\hline
$T_{\rm eff}$ (K) & $8350 \pm 140$ & 8409 \\
Radius ($R_\odot$) & $3.04 \pm 0.15$ & 3.27 \\
Luminosity ($L_\odot$) & $47.7 \pm 4.8$ & 47.8 \\
$\log g$ & $3.66 \pm 0.07$ & 3.75 \\
$P_{\rm orb}$ (days) & $2710 \pm 60$ & 2708 \\
\hline
\end{tabular}
\label{Table2}
\end{table}

\subsection{Evolution of the Donor and White Dwarf Formation}

\begin{figure*}[h]
\centering
\begin{subfigure}{0.5\textwidth}
\centering
\includegraphics[width=\linewidth]{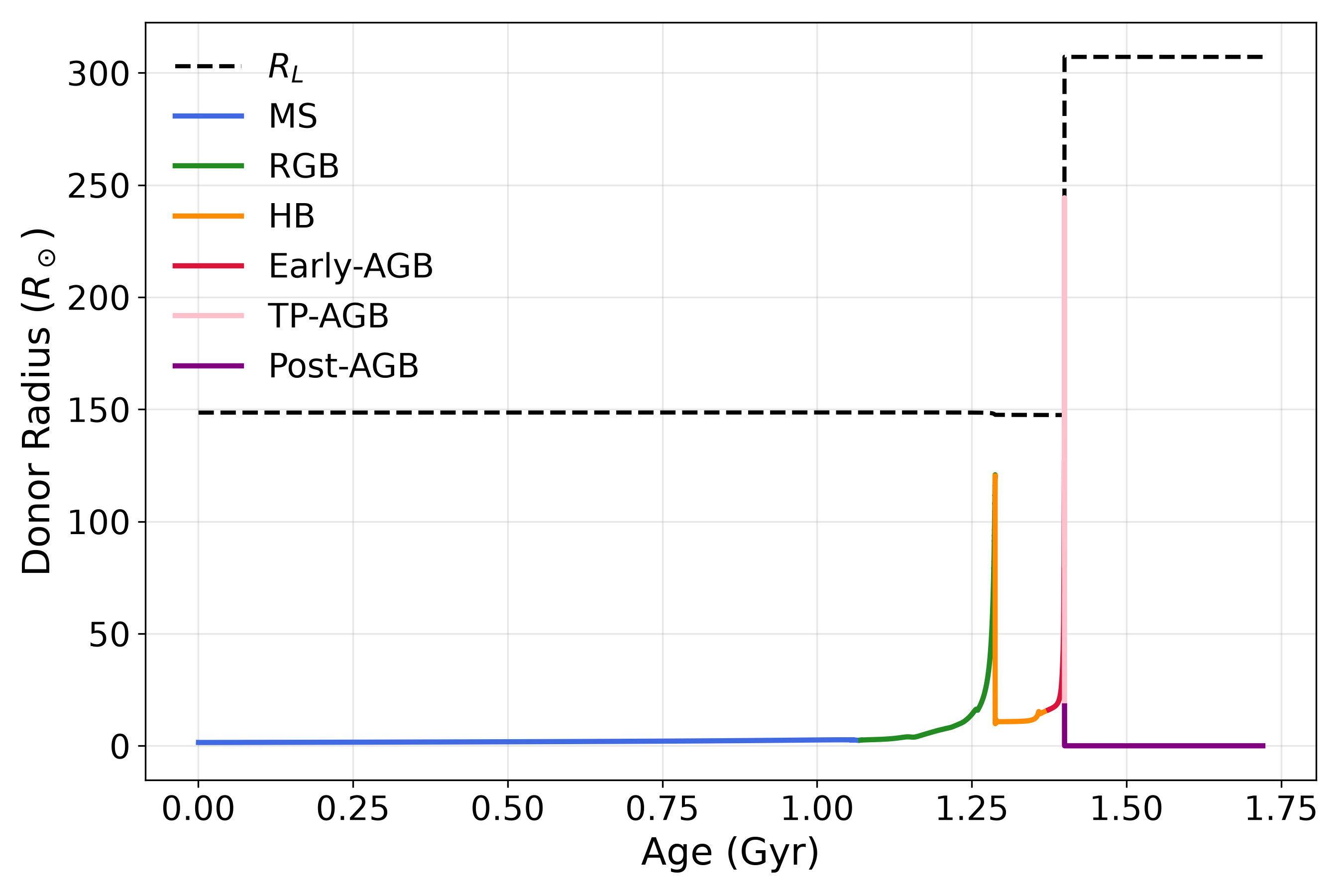}
\caption{}
\label{Figure2a}
\end{subfigure}
%\hfill
%\begin{subfigure}{0.4\textwidth}
%\centering
%\includegraphics[width=\linewidth]{Figure2b.png}
%\caption{}
%\label{Figure2b}
%\end{subfigure}
\hfill
\begin{subfigure}{0.5\textwidth}
\centering
\includegraphics[width=\linewidth]{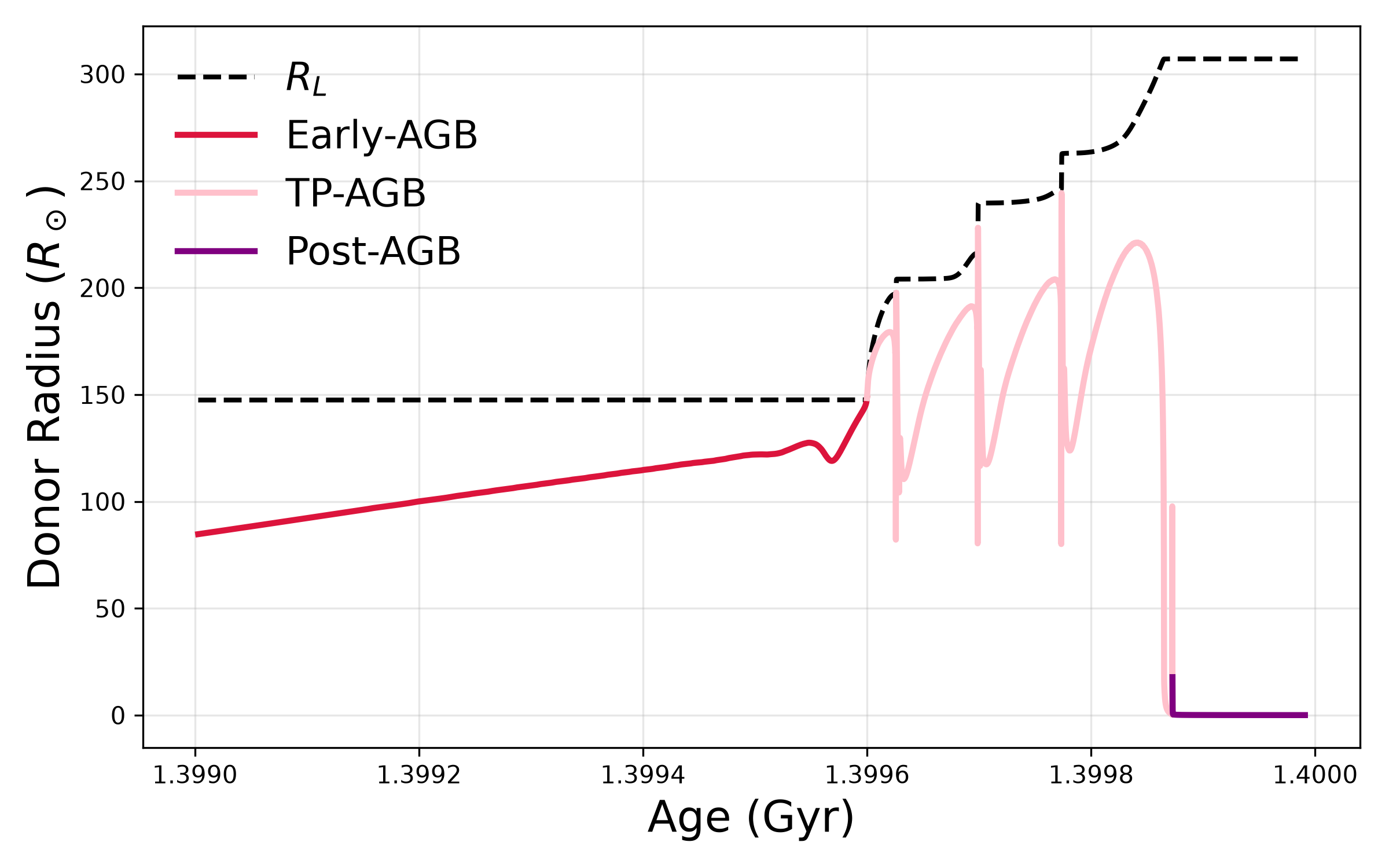}
\caption{}
\label{Figure2c}
\end{subfigure}

\caption{Evolution of the donor radius relative to the Roche-lobe radius. (a) Age–radius evolution of the donor star. (b) Zoomed view of the TP--AGB phase illustrating large-amplitude radius oscillations associated with thermal pulses.}
\label{Figure2}
\end{figure*}

\begin{figure*}[t]
\centering
\begin{subfigure}{0.4\textwidth}
\centering
\includegraphics[width=\linewidth]{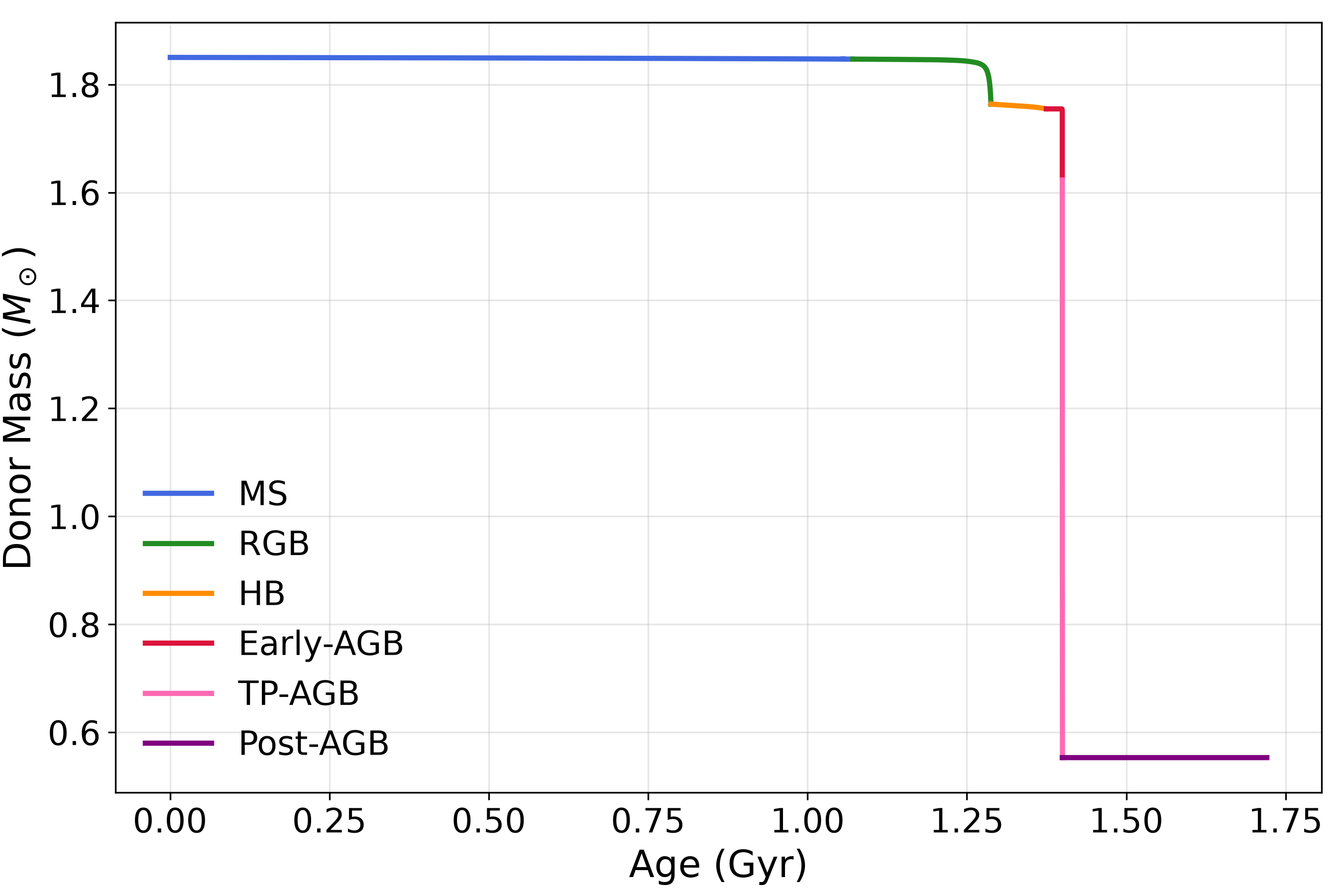}
\caption{}
\label{Figure3a}
\end{subfigure}
\hfill
\begin{subfigure}{0.4\textwidth}
\centering
\includegraphics[width=\linewidth]{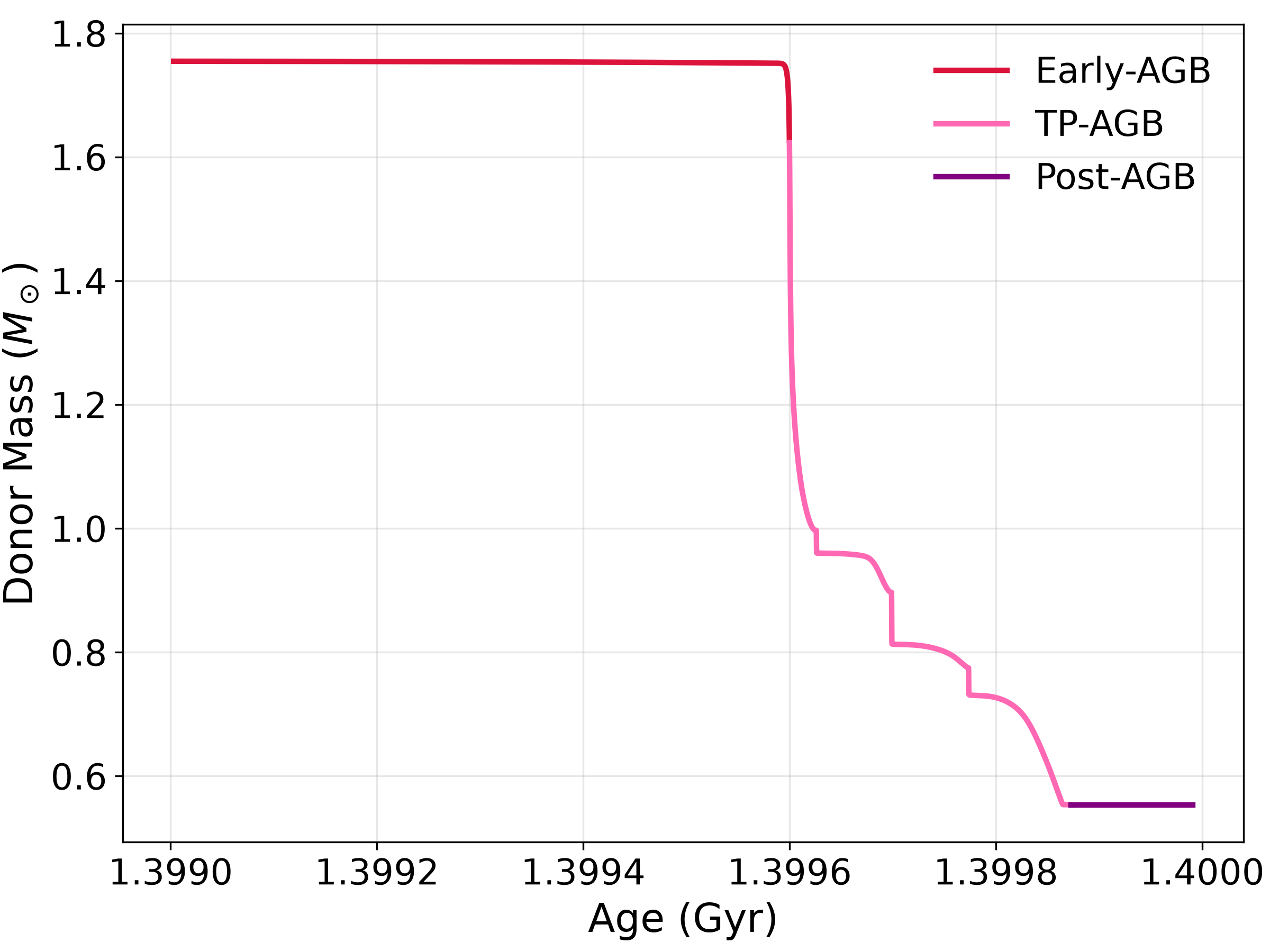}
\caption{}
\label{Figure3b}
\end{subfigure}
\caption{Donor mass evolution as a function of age. \textit{Left panel (Figure~\ref{Figure3a}):} Full evolutionary track showing the donor mass across different evolutionary phases (MS, RGB, HB, early-AGB, TP-AGB, and post-AGB). \textit{Right panel (Figure~\ref{Figure3b}):} Zoomed-in view of the late evolutionary stages (early-AGB, TP-AGB, and post-AGB), highlighting the detailed mass-loss behavior during the TP-AGB phase.}
\label{Figure3}
\end{figure*}

The evolution of the donor star is marked by the growth of its stellar radius relative to the Roche-lobe radius, which determines the dominant mode of mass transfer throughout the binary history. The overall evolution of the donor radius is shown in Figure \ref{Figure2a}, while Figure \ref{Figure2c} present zoomed view of the TP–AGB phases. The corresponding donor mass evolution is shown in Figure \ref{Figure3a}, with a zoomed view of the TP–AGB phase in Figure \ref{Figure3b}. Initially the system consists of a $1.85 M_{\odot}$ donor and a less massive ($1.67 M_{\odot}$) main-sequence companion in a wide binary. As illustrated in Figure \ref{Figure2a}, the Roche-lobe radius remains much larger than the stellar radius throughout the early evolution. During the main-sequence phase the donor radius remains small ($R_d \sim 1.5 R_{\odot}$) and nearly constant, while the Roche-lobe radius is $\sim 148 R_{\odot}$. Consequently the system remains fully detached and no significant mass transfer occurs. At the end of the main-sequence phase ($t = 1.06$ Gyr) the donor mass is $M_d = 1.84 M_{\odot}$ (Figure \ref{Figure3a}), confirming that negligible mass loss has taken place during this stage. After hydrogen exhaustion the donor ascends the RGB. During this phase the stellar radius increases smoothly and monotonically while remaining well below the Roche-lobe radius. Because $R_d < R_{\rm L,d}$ at all times, the binary remains detached and interaction occurs only through stellar winds. The donor mass remains nearly constant during this stage ($M_d = 1.84 M_{\odot}$ at $t = 1.07$ Gyr), indicating that mass loss during the early RGB evolution is negligible. The gradual expansion of the envelope reduces the separation between the stellar surface and the Roche lobe, allowing weak wind mass transfer to begin.

At $t = 1.28$ Gyr, the donor ignites helium in its core and enters the HB phase. The stellar radius reaches a first major peak at the tip of the RGB and subsequently undergoes a rapid contraction associated with the structural reconfiguration of the star following helium ignition. After core helium exhaustion the star evolves to the early-AGB. During this phase the envelope expands again and the stellar radius increases significantly, approaching the Roche-lobe boundary. At the beginning of the early-AGB phase ($t = 1.37$ Gyr) the donor mass is  1.75 $M_{\odot}$. Although the stellar radius still remains within the Roche lobe ($R_d < R_{\rm L,d}$), the large and extended envelope allows the stellar wind to become gravitationally focused toward the companion, enabling efficient WRLOF. The most significant evolution occurs during the TP–AGB phase. As the donor loses mass through WRLOF and episodic RLOF, the binary orbit evolves continuously, causing the Roche-lobe radius to change throughout this phase. As shown in the zoomed view in Figure \ref{Figure2c}, the donor radius is comparable to the Roche lobe radius (148 R$_{\odot}$) at 1.4 Gyr. This configuration marks the onset of classical RLOF. During subsequent thermal pulses the stellar radius exhibits large-amplitude oscillations and can exceed the $R_{\rm L,d}$, reaching values up to  223.53 $R_{\odot}$. These oscillations correspond to repeated expansion–contraction cycles of the envelope and produce short episodes of direct RLOF. The mass decreases rapidly during this phase (Figure \ref{Figure3a}). In particular, the donor mass declines from $\sim 1.6 M_{\odot}$ to $\sim 0.5 M_{\odot}$ as the envelope is removed through enhanced mass loss driven by the thermal pulses (Figure \ref{Figure3b}). Following the termination of the TP–AGB phase the envelope is rapidly expelled and the radius of the remaining donor contracts dramatically. At $t \sim 1.4$ Gyr the star enters the post-AGB stage with a mass of 0.55 $M_{\odot}$ and the stellar radius decreases to $\sim 17.8 R_{\odot}$.  At 17.8 R$_{\odot}$ there is still substantial remaining non-degenerate material that need to be removed, leaving behind the degenerate core at the WD radius. Eventually, the radius decreases to $\sim 0.01 R_{\odot}$ as the star cools along the WD track by $t \sim 1.72$ Gyr. During this stage the Roche-lobe radius expands to $\sim 307 R_{\odot}$. Because $R \ll R_{\rm L}$ from this point onward, mass transfer ends and the system evolves as a detached binary consisting of a BSS and a WD companion. The properties of the WD companion obtained from the best-fit model are summarized in Table \ref{Table3}.

\begin{table}[t]
\centering
\caption{Properties of the WD companion from the best-fit model.}
\begin{tabular}{lc}
\hline
Parameter & Value \\
\hline
Mass ($M_\odot$) & 0.55 \\
Effective temperature, $T_{\rm eff}$ (K) & 12520 \\
Luminosity ($L_\odot$) & 0.004 \\
Radius ($R_\odot$) & 0.013 \\
Surface gravity, $\log g$ & 7.92 \\
\hline
\end{tabular}
\label{Table3}
\end{table}

\subsection{Evolution of the Accretor and BSS Formation}

\begin{figure*}[h]
\centering
\begin{subfigure}{0.4\textwidth}
\centering
\includegraphics[width=\linewidth]{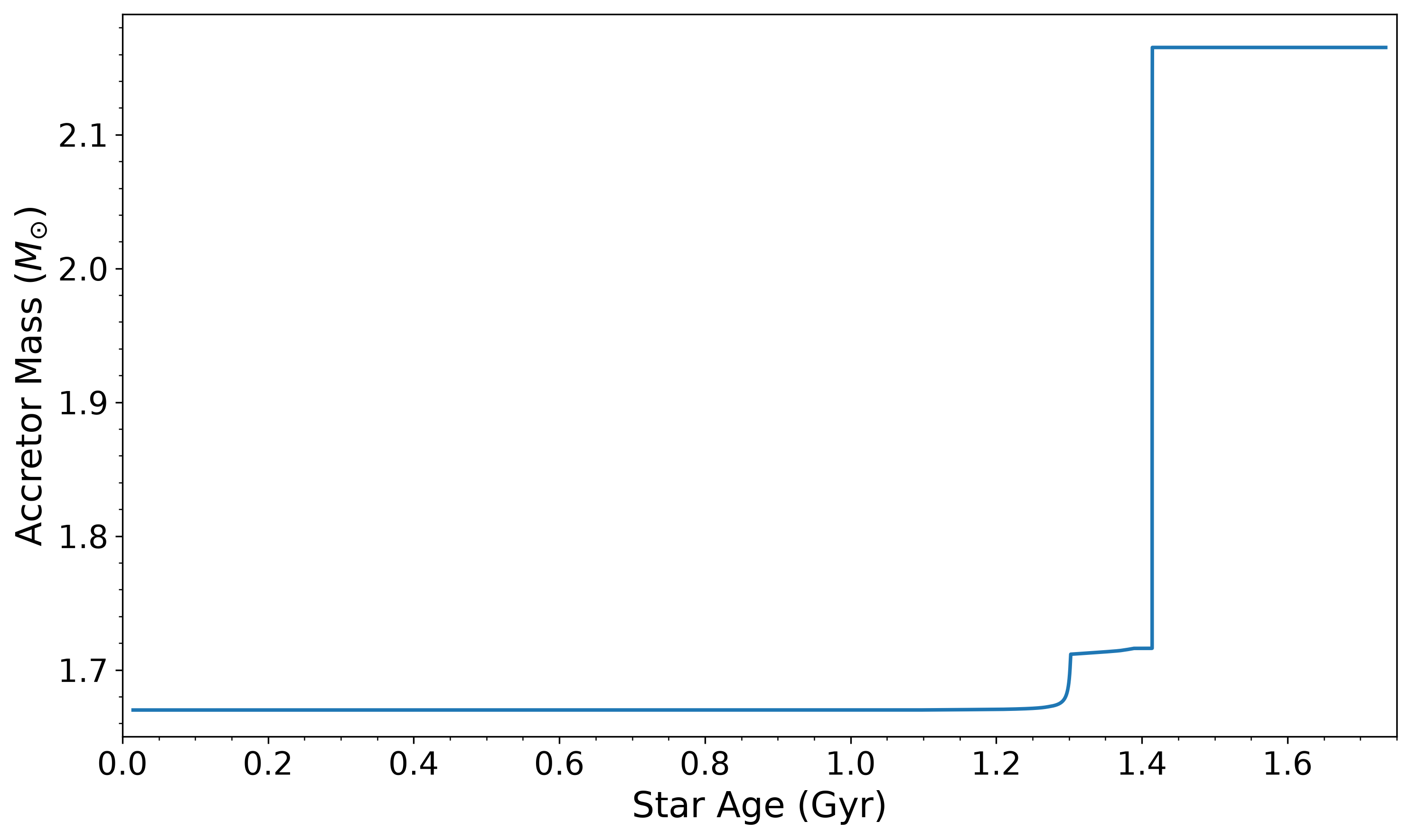}
\caption{}
\label{Figure4a}
\end{subfigure}
\hfill
\begin{subfigure}{0.4\textwidth}
\centering
\includegraphics[width=\linewidth]{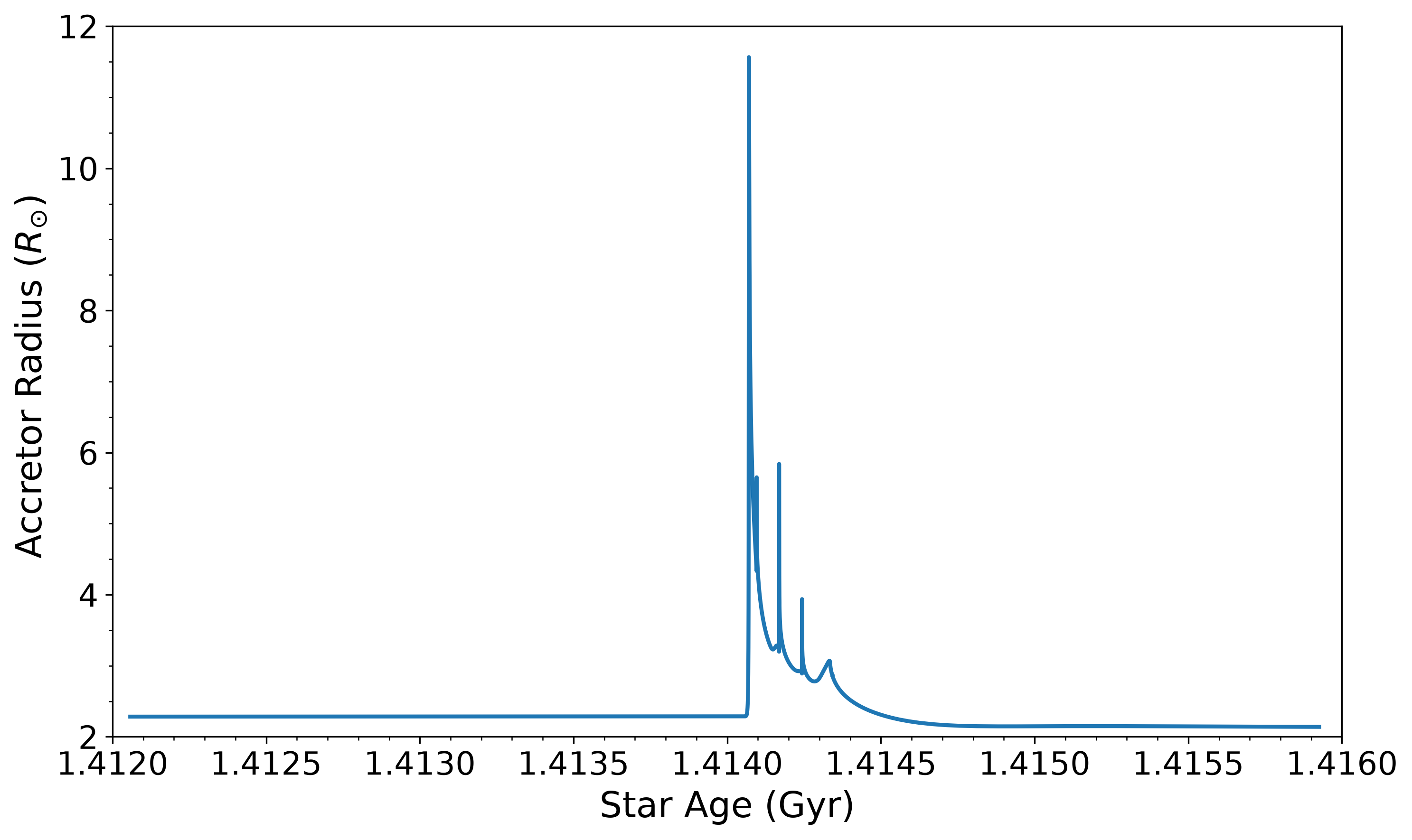}
\caption{}
\label{Figure4b}
\end{subfigure}
\caption{Left panel: Evolution of the accretor mass as a function of stellar age obtained from the best-fit MESA binary evolution model. Right panel: Zoomed view of the accretor radius during the rapid mass-transfer phase.}
\label{Figure4}
\end{figure*}

The evolution of the accretor is driven by mass transfer from the donor and ultimately leads to the formation of the BSS. Figure~\ref{Figure4a} shows the evolution of the accretor mass as a function of stellar age, while Figure~\ref{Figure4b} presents a zoomed view of the accretor radius during the rapid mass-transfer episode. At the beginning of the modeled binary evolution, the secondary star is a main-sequence companion with a mass (M$_{a}$) of 1.670 $M_{\odot}$ and a radius (R$_{a}$) of 1.45 $R_{\odot}$ at an age of  0.01 Gyr. During the early evolution, the donor remains well inside its Roche lobe and the binary stays in a detached configuration. Consequently, the accretor properties remain essentially unchanged and its mass stays close to the initial value. The first evidence of interaction appears once the donor begins to lose material through stellar winds. By 1.27 Gyr, the accretor mass has increased slightly to 1.672 $M_{\odot}$, while its radius has expanded to 2.27 $R_{\odot}$, primarily as a result of normal stellar evolution. During this stage, the donor remains within its Roche lobe, and the small increase in the accretor mass occurs entirely through wind capture. Because only a small fraction of the donor wind is accreted, the corresponding change in the donor mass is difficult to distinguish in the full donor mass-evolution plot. As the donor evolves through the subsequent core-helium-burning phase, wind accretion continues and the accretor gains mass gradually. By 1.29 Gyr, the accretor mass reaches 1.679 $M_{\odot}$, accompanied by an increase in radius to 2.30 $R_{\odot}$.

A much stronger structural response occurs later, when the donor evolves through the AGB and TP-AGB phases and the mass-transfer rate increases through efficient WRLOF and RLOF. By 1.41 Gyr, the accretor mass has increased to 1.75 $M_{\odot}$, while its radius has expanded to 3.94 $R_{\odot}$. Shortly afterward, during the rapid mass-transfer episode, the accretor mass grows to 1.84 $M_{\odot}$, and its radius reaches a maximum value of 11.57 $R_{\odot}$, as shown in Figure~\ref{Figure4b}. Following this brief expansion, the accretor begins to thermally readjust while mass transfer continues. Its mass increases further to 1.89 $M_{\odot}$, while the radius decreases to 8.95 $R_{\odot}$, indicating the onset of structural relaxation after the peak of the mass-transfer episode. As further material is transferred from the donor envelope, the accretor mass grows to 2.00 $M_{\odot}$, while its radius contracts significantly to  3.25 $R_{\odot}$. After the strongest mass-transfer episode ends and the donor begins its transition toward the WD stage, the transfer rate declines substantially and the accretor gradually approaches thermal equilibrium. The accretor continues to gain a smaller amount of mass through the wind mass-transfer and by 1.73 Gyr, it reaches a mass of 2.16 $M_{\odot}$ with a radius of 3.70 $R_{\odot}$. The binary evolution ends with the formation of a BSS of mass 2.17 M$_{\odot}$ and a WD companion. 

\subsection{Central Hydrogen and Helium Abundance Evolution of the BSS and WD}

\begin{figure*}[h]
\centering
\begin{subfigure}{0.48\textwidth}
\centering
\includegraphics[width=\linewidth]{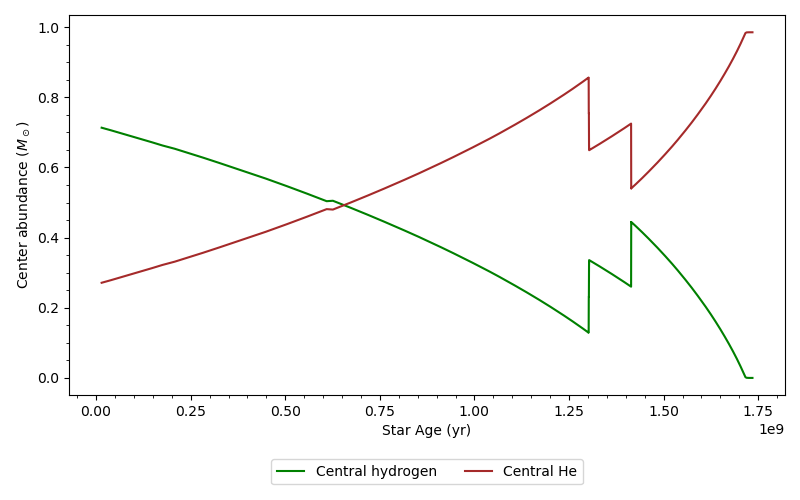}
\caption{}
\label{Figure5a}
\end{subfigure}
\hfill
\begin{subfigure}{0.48\textwidth}
\centering
\includegraphics[width=\linewidth]{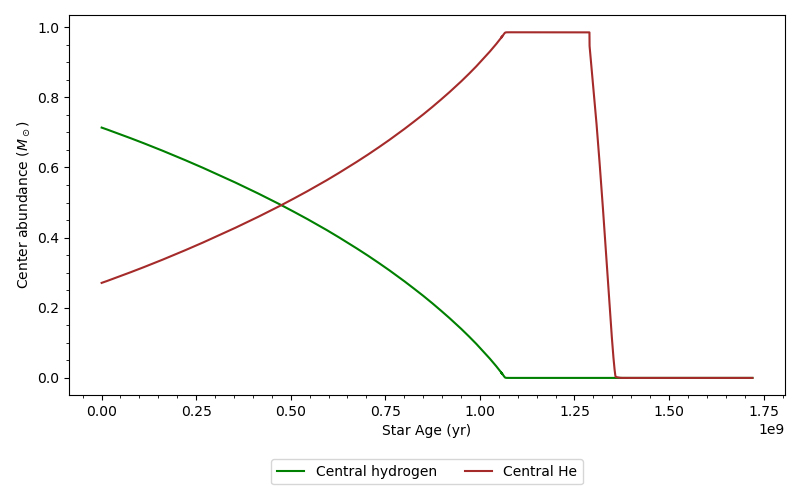}
\caption{}
\label{Figure5b}
\end{subfigure}

\caption{Central hydrogen ($X_{\rm c}$) and helium ($Y_{\rm c}$) abundances as a function of stellar age. (a) Accretor (BSS progenitor): shows core rejuvenation due to mass accretion during WRLOF and RLOF phases. (b) Donor (WD progenitor): shows standard stellar evolution followed by envelope stripping and WD formation with negligible central hydrogen and helium.}
\label{Figure5}

\end{figure*}

Figure \ref{Figure5a} shows the temporal evolution of the central hydrogen ($X_{\rm c,a}$) and helium ($Y_{\rm c,a}$) abundances of the accretor/BSS progenitor as a function of stellar age. During the early evolution (t = 1.085 Gyr), the star follows a normal main-sequence evolutionary path dominated by core hydrogen burning. In this phase, the central hydrogen abundance decreases from $X_{\rm c,a} = 0.71$ to $X_{\rm c,a} = 0.28$, while the helium abundance increases from $Y_{\rm c,a} = 0.27$ to $Y_{\rm c,a} = 0.71$. The onset of binary interaction occurs through weak wind accretion ($t = 1.086$–$1.294$ Gyr). As very little mass is deposited, the internal nuclear evolution remains largely unaffected, and the central hydrogen abundance continues to decline from $X_{\rm c,a} = 0.28$ to $X_{\rm c,a} = 0.13$, while helium increases to $Y_{\rm c,a} = 0.80$. A significant transition occurs when the system enters the WRLOF regime ($t = 1.294$–$1.414$\,Gyr), during which a substantial amount of hydrogen-rich material is accreted. This leads to a reversal in the abundance trends: the central hydrogen abundance increases from $X_{\rm c,a} = 0.13$ to $X_{\rm c,a} = 0.31$, while the helium abundance decreases from $Y_{\rm c,a} = 0.85$ to $Y_{\rm c,a} = 0.68$. The rejuvenation becomes most pronounced during the brief classical RLOF phase around $t = 1.414$\,Gyr, when the mass-transfer rate reaches its maximum. During this stage, the central hydrogen abundance further increases from $X_{\rm c,a} = 0.31$ to $X_{\rm c,a} = 0.39$, while the helium abundance decreases from $Y_{\rm c,a} = 0.68$ to $Y_{\rm c,a} = 0.60$. Following RLOF, the system enters a post-RLOF wind-accretion phase, during which accretion continues at a reduced efficiency due to orbital widening. In this stage, the central hydrogen abundance increases gradually to $X_{\rm c,a} = 0.44$, while helium decreases to $Y_{\rm c,a} = 0.54$, indicating moderate continued rejuvenation. After mass transfer ceases, the accretor evolves as an isolated star. The central hydrogen abundance then decreases from $X_{\rm c,a} = 0.44$ to nearly zero, while helium increases to $Y_{\rm c,a} = 0.99$, marking the eventual exhaustion of core hydrogen in the BSS and its subsequent evolution off the main sequence.

Figure \ref{Figure5b} shows the temporal evolution of the central hydrogen ($X_{\rm c,d}$) and helium ($Y_{\rm c,d}$) abundances of the donor star, which ultimately evolves into the WD companion. At the beginning of the evolution, the donor is a normal main-sequence star with $X_{\rm c,d} = 0.71$ and $Y_{\rm c,d} = 0.27$. During the main-sequence phase, steady core hydrogen burning leads to a gradual decrease in hydrogen and a corresponding increase in helium. By the end of the main sequence ($t = 1.07$\,Gyr), the central hydrogen abundance has declined to nearly zero ($X_{\rm c,d} = 0$), while helium has increased to $Y_{\rm c,d} = 0.99$, indicating the exhaustion of core hydrogen. The star then evolves through the RGB, during which the core remains hydrogen-depleted and helium-rich. Between $t = 1.07$\,Gyr and $t = 1.29$\,Gyr, the central abundances remain approximately constant ($X_{\rm c,d} = 0$, $Y_{\rm c,d} = 0.99$), reflecting an inert helium core. Helium ignition occurs near the RGB tip ($t = 1.287$\,Gyr), after which the star enters the core helium-burning phase. During this stage, helium is progressively converted into heavier elements, leading to a rapid decline in $Y_{\rm c,d}$ to $\sim 10^{-4}$ by $t = 1.37$\,Gyr, while hydrogen remains absent in the core. Subsequently, the donor evolves onto the AGB and enters the thermally pulsing AGB phase. During this late evolutionary stage, strong mass loss driven by stellar winds and binary interaction removes the envelope of the star. As a result, the donor evolves into a compact remnant dominated by its degenerate core, and both central hydrogen and helium abundances become effectively negligible ($X_{\rm c,d} = 0$, $Y_{\rm c,d} \sim 10^{-30}$). By the end of the simulation ($t = 1.72$\,Gyr), the donor has evolved into a compact WD with negligible central hydrogen and helium abundances, consistent with a fully stripped, degenerate remnant.

\subsection{Orbital Period Evolution and Binary Mass Exchange}

\begin{figure*}[t]
\centering
\begin{subfigure}{0.7\textwidth}
\centering
\includegraphics[width=\linewidth]{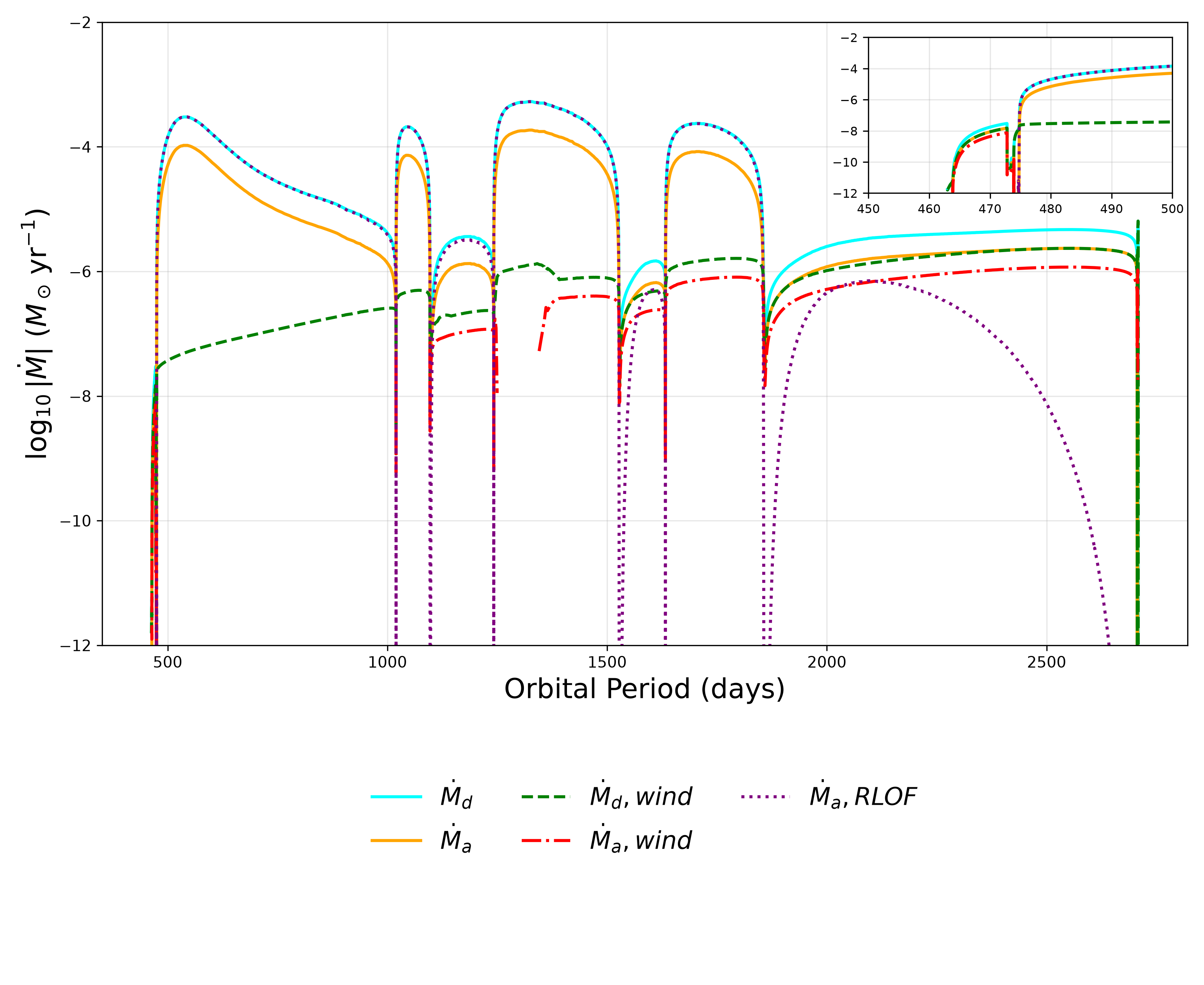}
\caption{}
\label{Figure6a}
\end{subfigure}
\hfill
\begin{subfigure}{0.70\textwidth}
\centering
\includegraphics[width=\linewidth]{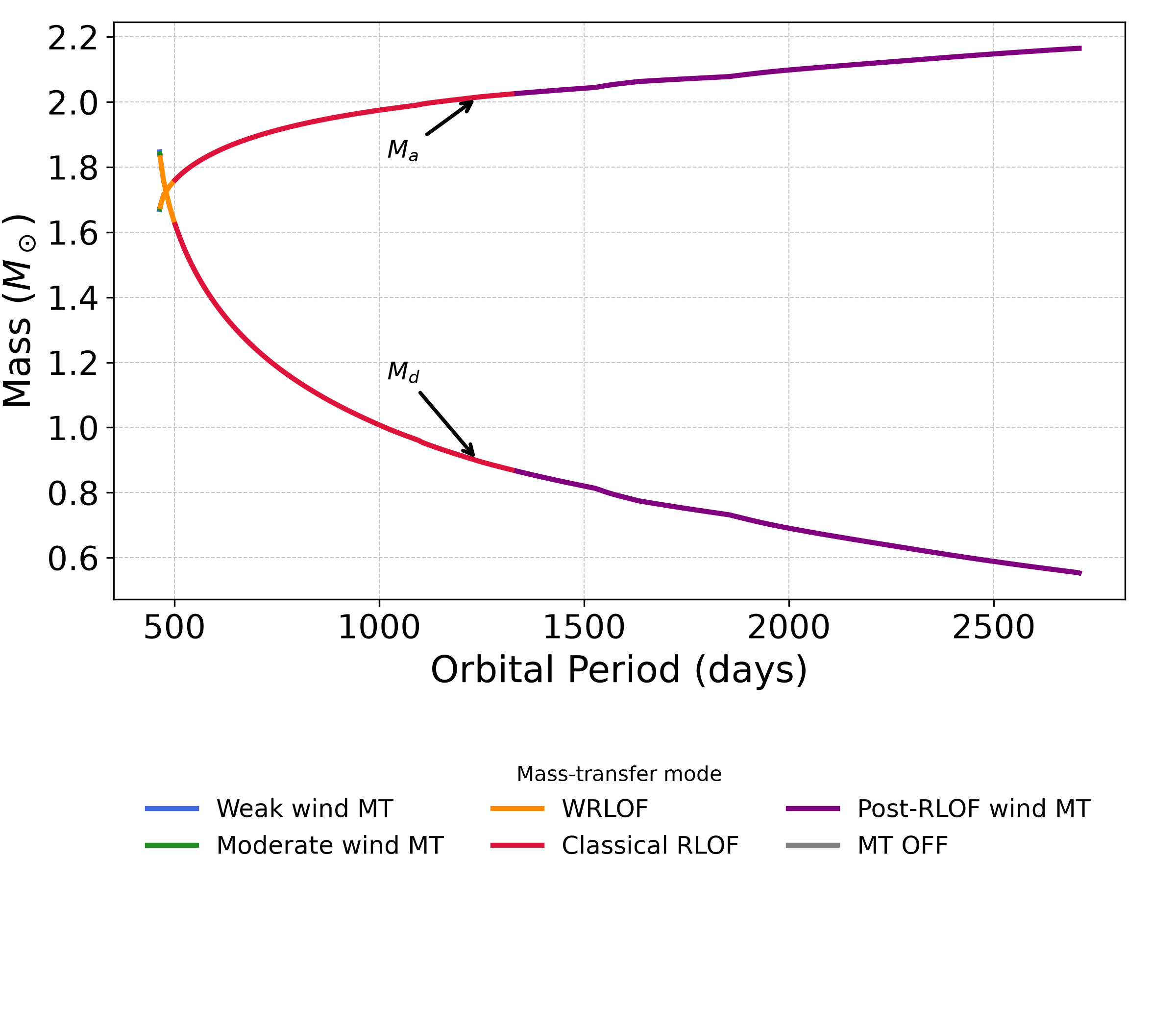}
\caption{}
\label{Figure6b}
\end{subfigure}

\caption{Evolution of mass transfer and stellar masses as a function of orbital period. (a) Mass-transfer rates, including the donor total mass-loss rate ($\dot{M}_d$), donor wind mass-loss rate ($\dot{M}_{d,\mathrm{wind}}$), accretor total mass-gain rate ($\dot{M}_a$), wind accretion rate ($\dot{M}_{a,\mathrm{wind}}$), and RLOF accretion rate ($\dot{M}_{a,\mathrm{RLOF}}$). The inset highlights the detail of when wind mass transfer dominates and the early onset of RLOF at short orbital periods. (b) Evolution of the stellar masses, showing the decrease of the donor mass ($M_d$) and the corresponding growth of the accretor mass ($M_a$), leading to the formation of a BSS.}
\label{Figure6}

\end{figure*}

The orbital evolution of an interacting binary is governed by changes in the stellar masses, the total system mass, and the orbital angular momentum resulting from mass transfer and mass loss \citep{tauris2006formation}. The relative importance of these processes determines whether the orbital separation and orbital period increase or decrease during binary evolution. Figures~\ref{Figure6a} and \ref{Figure6b} illustrate the evolution of the mass-transfer rates and stellar masses, respectively, with respect to the orbital period.

The total mass-loss rate from the donor consists of contributions from stellar winds and RLOF
\begin{equation}
\dot{M}_d = \dot{M}_{d,\mathrm{wind}} + \dot{M}_{d,\mathrm{RLOF}},
\end{equation}
while the total mass accretion rate onto the accretor is
\begin{equation}
\dot{M}_a = \dot{M}_{a,\mathrm{wind}} + \dot{M}_{a,\mathrm{RLOF}}.
\end{equation}

The wind-accretion component is computed as
\begin{equation}
\dot{M}_{\rm a,wind} = \beta_{\rm acc}\,\dot{M}_{\rm d,wind},
\end{equation}
where $\beta_{\rm acc}$ represents the wind-accretion efficiency. During the RGB phase, we adopt a constant value of $\beta_{\rm acc}=0.5$, while during the AGB phase it is computed using the WRLOF prescription of \citet{abate2013wind}, following the implementation of \citet{sun2023wocs}. The accretion rate during RLOF is parameterized as
\begin{equation}
\dot{M}_{a,\mathrm{RLOF}} = -\beta\,\dot{M}_{d,\mathrm{RLOF}},
\end{equation}
 where  $\beta$ represents the RLOF mass-accretion efficiency.

During the main-sequence phase of the donor ($P \sim 463.00$--$463.83\,\mathrm{days}$), the binary remains detached and no mass transfer occurs. The donor loses a negligible amount of mass through a weak stellar wind ($\log \dot{M}_d \sim -11.8$ to $-11.2$; inset in Figure~\ref{Figure6a}), but this material escapes the system and is not accreted by the companion. Both stellar masses remain essentially constant during this period (Figure~\ref{Figure6b}), and the orbital period changes negligibly. Wind mass transfer begins as the donor ascends the RGB. During the initial weak-wind phase ($P = 463.83$--$464.35\,\mathrm{days}$), the donor mass decreases slightly from $1.846\,M_\odot$ to $1.841\,M_\odot$ ($\Delta M_d = -0.005\,M_\odot$), while the accretor gains $0.002\,M_\odot$ ($1.670$--$1.672\,M_\odot$). The mass-loss rate remains low ($\log \dot{M}_d \sim -11.3$ to $-9.7$), resulting in only a small amount of wind accretion and negligible orbital evolution. This indicates that the specific angular momentum carried away by the weak wind is insufficient to drive substantial changes in the binary orbit.

As the donor continues to evolve up the RGB, the stellar wind strengthens and the system enters a moderate wind mass-transfer regime ($P = 464.35$--$465.55\,\mathrm{days}$). During this phase, the donor mass decreases from $1.841\,M_\odot$ to $1.828,M_\odot$ ($\Delta M_d = -0.012\,M_\odot$), while the companion accretes only $0.006,M_\odot$. The increasing wind mass-loss rate ($\log \dot{M}_d \sim -9.7$ to $-8.7$) leads to a gradual expansion of the orbit. This orbital widening results from the combined effects of non-conservative mass transfer, mass loss from the system, and the changing binary mass ratio. A major transition occurs as the donor evolves through the HB and into the early AGB, where the system enters the WRLOF regime ($P = 465.55$--$500.95\,\mathrm{days}$). In this configuration, the slow AGB wind is gravitationally confined within the Roche lobe and preferentially directed toward the companion. Consequently, the donor loses $0.201\,M_\odot$ ($1.828 \rightarrow 1.627\,M_\odot$), while the companion accretes $0.080\,M_\odot$ ($1.679 \rightarrow 1.759\,M_\odot$). The mass-loss rates increase substantially, reaching values up to $\log \dot{M}_d \sim -3.8$. Compared with the preceding wind-accretion phase, WRLOF produces a more focused mode of mass transfer and a relatively larger amount of material accreted by the companion.

RLOF begins at $P \sim 475$ days, where $\dot{M}_{\mathrm{RLOF}}$ first becomes non-zero (inset of Figure~\ref{Figure6a}). As the donor evolves along the TP--AGB, the RLOF mass-transfer rate increases rapidly, and RLOF becomes the dominant mass-transfer channel ($P = 501.96$--$1328.84$ days). During this phase, the mass-transfer rate reaches $\log \dot{M} \sim -3.3$ (Figure~\ref{Figure6a}), leading to rapid removal of the donor envelope and continued mass accretion by the companion. The donor loses $0.754\,M_\odot$ ($1.623 \rightarrow 0.868\,M_\odot$), while the accretor gains $0.264\,M_\odot$ ($1.761 \rightarrow 2.025\,M_\odot$). This phase contributes the largest fraction of the total mass transferred to the companion (Figure~\ref{Figure6b}) and is accompanied by the most rapid increase in the orbital period. Following the termination of RLOF, mass transfer continues through stellar winds during the post-RLOF phase ($P = 1336.25$--$2708.44\,\mathrm{days}$). During this stage, the donor loses an additional $0.312\,M_\odot$ ($0.865 \rightarrow 0.553\,M_\odot$), while the accretor gains $0.138\,M_\odot$ ($2.026 \rightarrow 2.165\,M_\odot$). Once most of the remaining donor envelope is removed, the donor evolves into a compact WD remnant dominated by its degenerate core, with a final mass of $0.553\,M_\odot$. The orbital period reaches a final value of 2708.44 days. The overall mass-transfer properties of the system are summarized in Table~\ref{Table4}, which lists the total mass lost by the donor, the mass accreted by the companion, and the fraction of mass lost from the system.

\begin{table}[t]
\centering
\caption{Mass-transfer budget of the binary evolution from the best-fit model.}
\begin{tabular}{lc}
\hline
Quantity & Value  \\
\hline
Total mass lost by donor & 1.29 $M_\odot$\\
Total mass gained by accretor & 0.49 $M_\odot$ \\
Mass lost from system & 0.80 $M_\odot$ \\
Overall MT efficiency, $\eta_{\rm MT}$ & 0.38 \\
\hline
\end{tabular}
\label{Table4}
\end{table}

\subsection{Timescale Analysis of the Binary Interaction}
\subsubsection{Donor Radius Evolution and Mass-Loss Timescales}

\begin{figure*}[h]
\centering
\begin{subfigure}{0.48\textwidth}
\centering
\includegraphics[width=\linewidth]{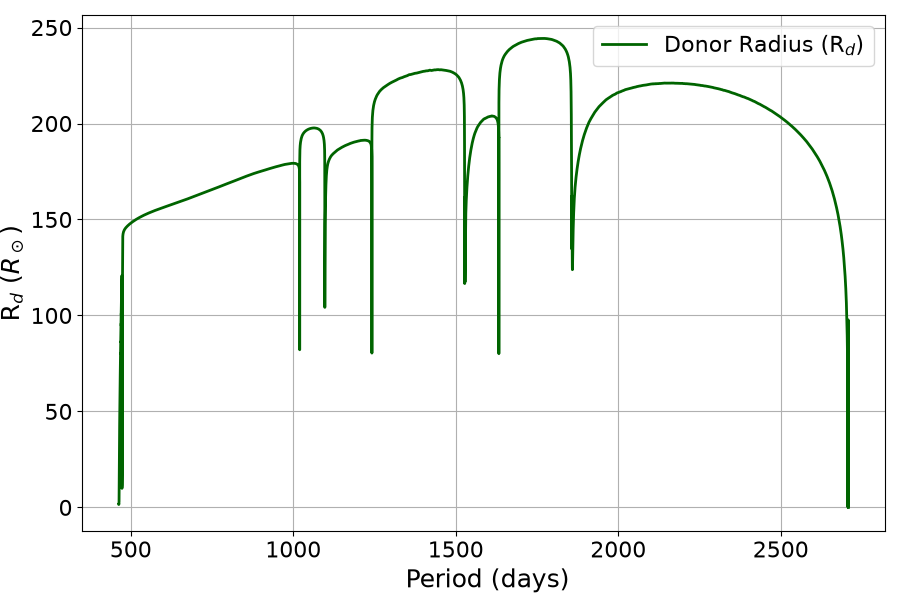}
\caption{}
\label{Figure7a}
\end{subfigure}
\hfill
\begin{subfigure}{0.48\textwidth}
\centering
\includegraphics[width=\linewidth]{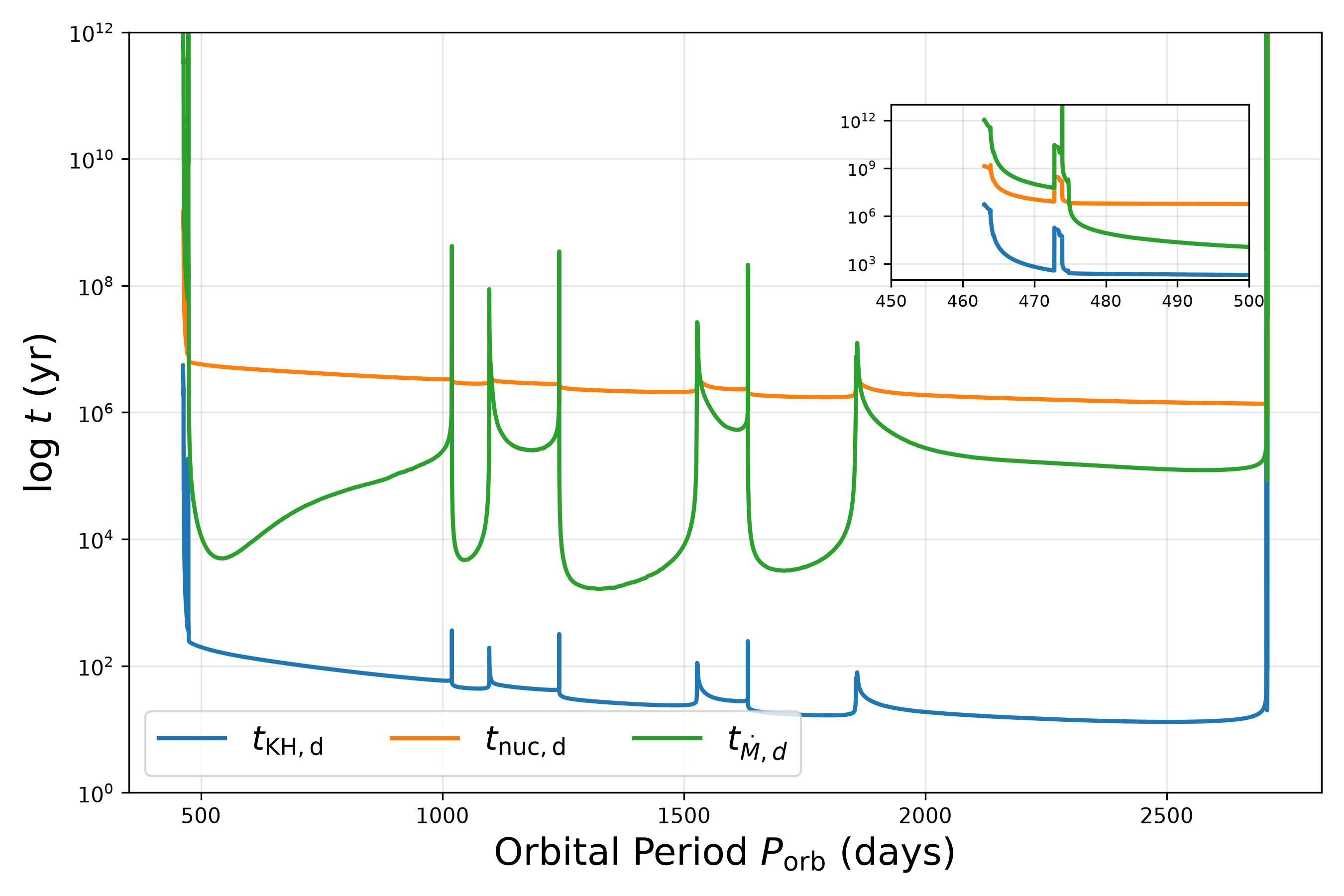}
\caption{}
\label{Figure7b}
\end{subfigure}
\caption{ Evolution of the donor star as a function of orbital period. (a) Donor radius ($R_d$) versus orbital period. (b) Corresponding evolution of the donor nuclear ($t_{\rm nuc,d}$), thermal ($t_{\rm KH,d}$), and mass-loss ($t_{\dot{M},d}$) timescales.}
\label{Figure7}

\end{figure*}

Figure \ref{Figure7a} shows the evolution of the donor radius as a function of orbital period, while Figure \ref{Figure7b} presents the corresponding nuclear ($t_{\rm nuc,d}$), thermal ($t_{\rm KH,d}$), and mass-loss ($t_{\dot{M},d}$) timescales.

The nuclear timescale is defined as
\begin{equation}
t_{\rm nuc} = 10^{10}\,{\rm yr}
\left(\frac{M/M_{\odot}}{L/L_{\odot}}\right),
\end{equation}
the thermal (Kelvin--Helmholtz) timescale as
\begin{equation}
t_{\rm KH} = \frac{G M^{2}}{R L},
\end{equation}
and the mass-loss timescale as
\begin{equation}
t_{\dot{M}} = \frac{M}{|\dot{M}|},
\end{equation}
where $M$, $R$, $L$, and $\dot{M}$ denote the instantaneous stellar
mass, radius, luminosity, and mass-loss rate, respectively.

At the beginning of the interaction, when the orbital period is $P_{\rm orb} = 463$–465 days, the donor is on the RGB with $R_d$ = 2$–5$ R$_\odot$. As shown in Figure \ref{Figure7b}, the nuclear timescale is $t_{\rm nuc,d}\sim10^{9}$ yr, while the thermal timescale of the envelope is much shorter, $t_{\rm KH,d}\sim10^{5}$–$10^{6}$ yr. The mass-loss timescale is significantly longer, $t_{\dot{M},d}\gtrsim10^{11}$–$10^{12}$ yr, indicating that stellar winds have essentially no structural influence on the donor at this stage. Consequently, the star expands smoothly as it ascends the RGB, and the orbital period increases gradually from $\sim463$ days to $\sim500$ days while the radius rapidly grows to 140 $–$ 150 $R_\odot$. As the donor continues to evolve, the stellar radius increases further while the orbital period expands to 1000 days. At this stage the donor reaches radii of 170 $–$ 190$ R_\odot$ (Figure \ref{Figure7a}). The stellar wind reduces the mass-loss timescale to $t_{\dot{M},d}\sim10^{5}$–$10^{6}$ yr. Although this remains significantly longer than the thermal adjustment timescale of the envelope ($t_{\rm KH,d}\sim10^{2}$–$10^{3}$ yr), it is now several orders of magnitude shorter than in the earlier RGB phase. This indicates that the donor continues to maintain thermal equilibrium ($t_{\rm KH,d} \ll t_{\dot{M},d}$), but mass loss has become sufficiently strong to influence the envelope structure. During the TP-AGB phase, the donor radius becomes very large, reaching 200 $–$ 245 $R_\odot$ while the orbital period increases from 1200 to 1800 days. In this regime the stellar radius undergoes repeated expansion–contraction cycles driven by thermal pulses in the helium-burning shell, which appear as the large loops seen in the radius–period diagram. The timescale evolution reflects this behavior: the mass-loss timescale further decreases to $t_{\dot{M},d}\sim10^{4}$–$10^{5}$ yr, remaining longer than the thermal timescale ($t_{\rm KH,d}\sim10^{2}$–$10^{3}$ yr) but approaching it more closely than in earlier phases. This hierarchy ($t_{\rm KH,d} \ll t_{\dot{M},d} \ll t_{\rm nuc,d}$, with $t_{\rm nuc,d}\sim10^{6}$–$10^{7}$ yr) implies that the envelope can rapidly readjust to perturbations while simultaneously being affected by ongoing mass removal. Consequently, the envelope evolution during this phase is governed by the interplay between thermal adjustment and enhanced mass loss, rather than by nuclear evolution in the core alone. As mass and angular momentum are removed from the system, the orbital period continues to increase steadily. By the end of the TP-AGB phase the orbital period has increased to 2600 $–$ 2700 days. At this point the donor rapidly loses its remaining envelope and the system settles into its final configuration as a wide BSS–WD binary.

\subsubsection{Accretor Structural Response to Mass Transfer}

\begin{figure*}[h]
\centering
\begin{subfigure}{0.45\textwidth}
\centering
\includegraphics[width=\linewidth]{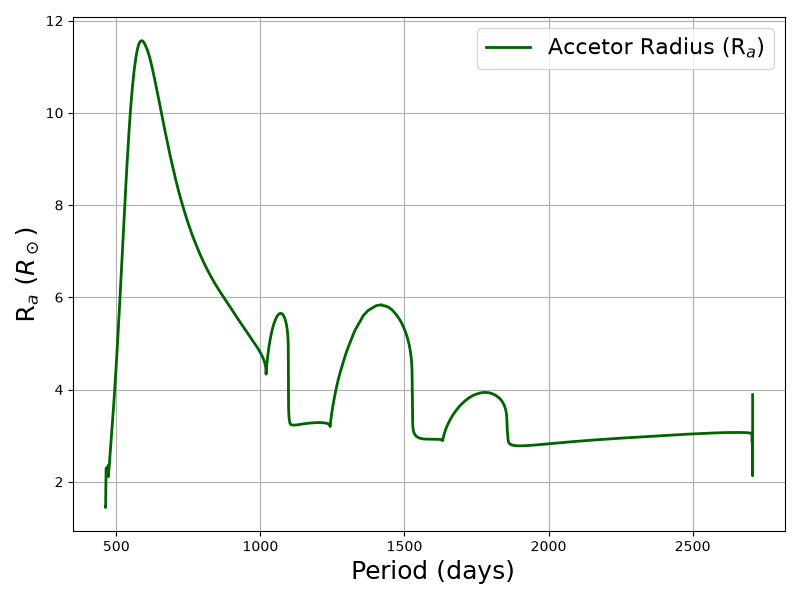}
\caption{}
\label{Figure8a}
\end{subfigure}
\hfill
\begin{subfigure}{0.5\textwidth}
\centering
\includegraphics[width=\linewidth]{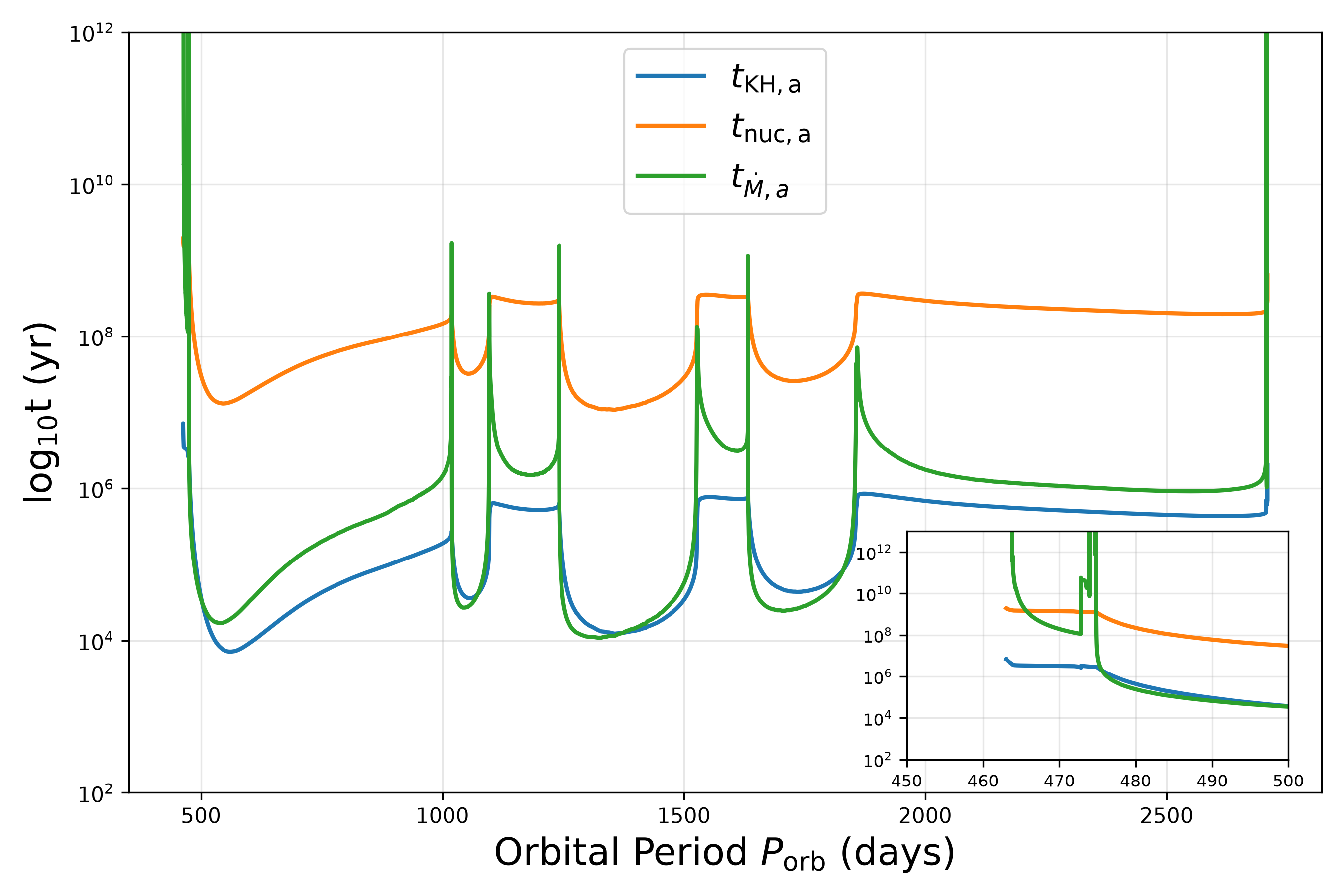}
\caption{}
\label{Figure8b}
\end{subfigure}
\caption{Evolution of the accretor as a function of orbital period.(a) Accretor radius ($R_a$) versus orbital period.(b) Corresponding evolution of the accretor nuclear ($t_{\rm nuc,a}$), thermal ($t_{\rm KH,a}$), and mass-gain ($t_{\dot{M},a}$) timescales.}
\label{Figure8}
\end{figure*}

The structural response of the accretor during the binary interaction can be understood by examining the evolution of the BSS radius together with the characteristic timescales that govern the accretion process. Figure \ref{Figure8a} shows the variation of the accretor radius as a function of orbital period, while Figure \ref{Figure8b} presents the corresponding evolution of the accretor mass-transfer timescale ($t_{\dot{M},a}$), Kelvin–Helmholtz timescale ($t_{\rm KH,a}$), and nuclear timescale ($t_{\rm nuc,a}$). The relative magnitudes of $t_{\dot{M},a}$ and $t_{\rm KH,a}$ determine how efficiently the accretor can thermally adjust to the incoming material.

The orbital period expands from $\sim500$ days to $\sim1000$ days while the accretor radius (R$_{a}$) grows from 2.5 $R_\odot$ to a maximum of 11.5 $R_\odot$. During this phase the mass-transfer timescale decreases to $t_{\dot{M},a}\sim10^{6}$–$10^{7}$ yr, approaching the thermal adjustment timescale of the accretor ($t_{\rm KH,a} \sim10^{4}$–$10^{5}$ yr). Because the accretor can only partially redistribute the accreted material on its thermal timescale, the stellar envelope expands significantly, producing the prominent peak in the accretor radius seen near $P_{\rm orb}\sim520$–600 days. At later times, as the orbital period increases further to 1000 $–$ 1500 days, the accretor radius settles to 3 $–$ 6 $R_\odot$. The mass-transfer timescale fluctuates between $t_{\dot{M},a}\sim10^{5}$ and $10^{7}$ yr due to episodic mass-transfer events associated with thermal pulses in the donor. These variations produce the smaller loops visible in the radius–period plane. During these phases $t_{\dot{M},a}$ becomes comparable to or shorter than the thermal timescale, causing temporary departures from thermal equilibrium and moderate expansions of the accretor envelope. The strongest structural response occurs during brief phases when the donor nearly fills its Roche lobe during the thermally pulsing AGB phase. In this regime the mass-transfer timescale drops to $t_{\dot{M},a}\lesssim10^{4}$–$10^{5}$ yr, becoming shorter than the thermal timescale ($t_{\rm KH,a}\sim10^{5}$–$10^{6}$ yr). Because the accretor cannot thermally relax fast enough, the stellar envelope expands rapidly, producing the sharp spikes visible in the radius–period diagram. As the donor loses its envelope and the interaction weakens, the mass-transfer rate declines and $t_{\dot{M},a}$ increases again to $\gtrsim10^{10}$ yr. The accretor radius gradually contracts and stabilizes at 3 $R_\odot$.

\subsection{Orbital Angular Momentum Evolution and Loss Mechanisms}

\begin{figure*}[h]
\centering
\begin{subfigure}{0.48\textwidth}
\centering
\includegraphics[width=\linewidth]{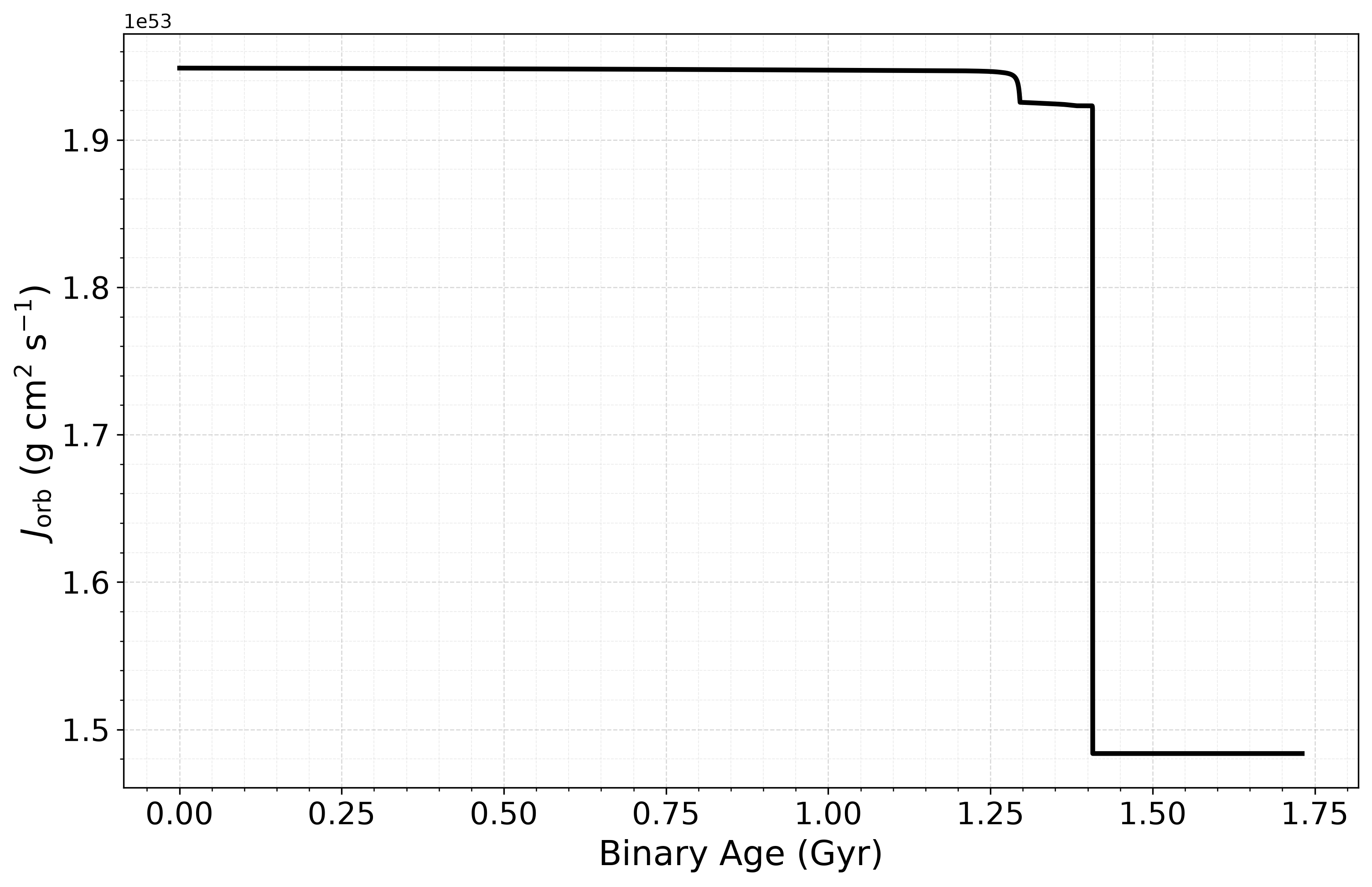}
\caption{}
\label{Figure9a}
\end{subfigure}
\hfill
\begin{subfigure}{0.48\textwidth}
\centering
\includegraphics[width=\linewidth]{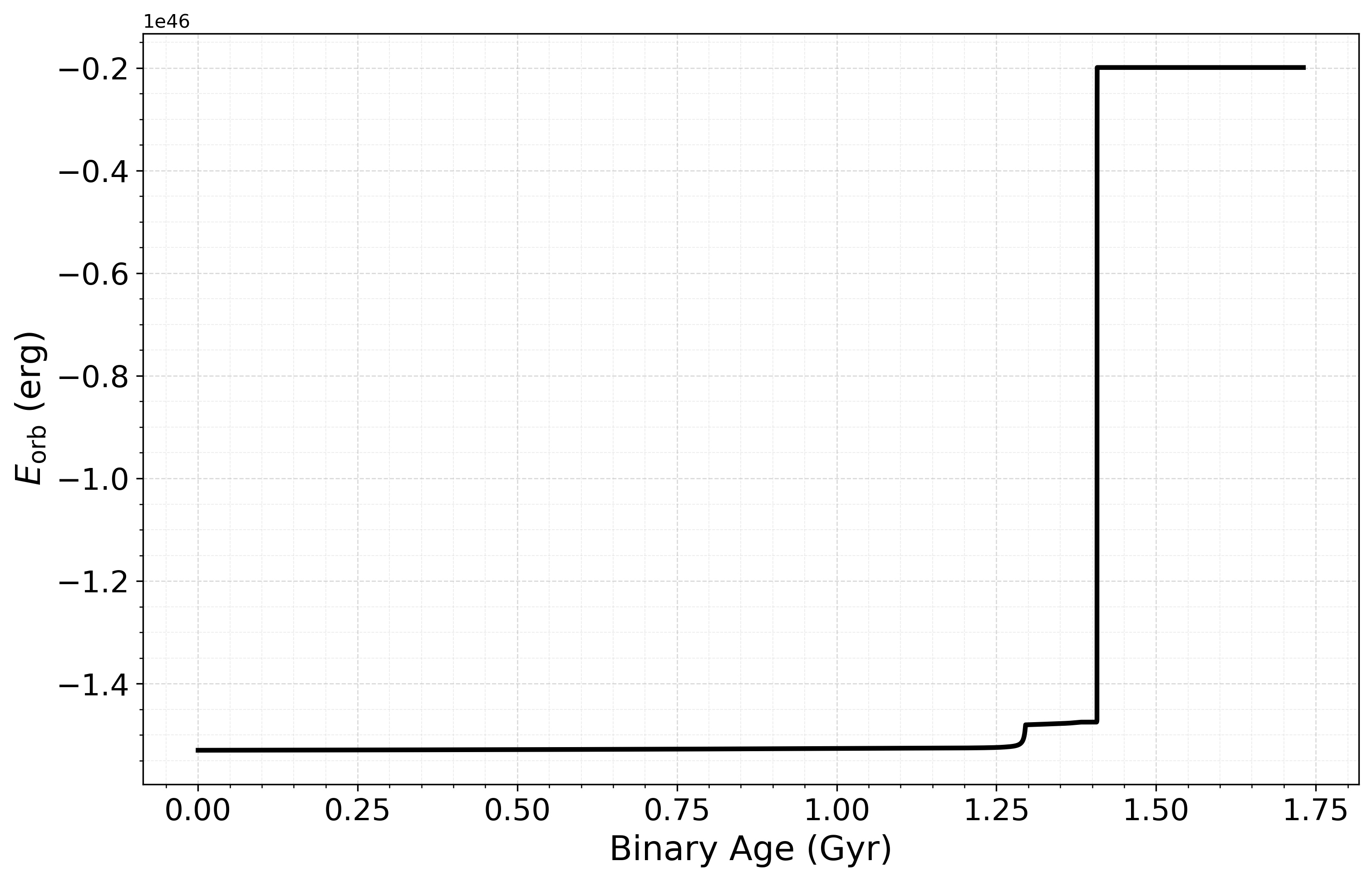}
\caption{}
\label{Figure9b}
\end{subfigure}
\caption{
Evolution of the orbital angular momentum (left panel) and orbital energy (right panel) as a function of binary age. The two panels together illustrate the loss of angular momentum and the corresponding change in the orbital energy during the mass-transfer phase.}
\label{Figure9}
\end{figure*}

The orbital angular momentum remains nearly constant during the MS, RGB, and core helium-burning phases, decreasing only slightly from $1.948\times10^{53}$ to $1.923\times10^{53}\,\mathrm{g\,cm^{2}\,s^{-1}}$. The most significant change occurs during the TP--AGB phase, when $J_{\rm orb}$ drops rapidly to $1.48\times10^{53}\,\mathrm{g\,cm^{2}\,s^{-1}}$ at $1.4\,\mathrm{Gyr}$ (Figure \ref{Figure9a}). Similarly, the orbital energy changes only modestly from $-1.52\times10^{46}$ to $-1.47\times10^{46}\,\mathrm{erg}$ prior to the TP--AGB phase, before increasing sharply (becoming less negative) to $-2.0\times10^{45}\,\mathrm{erg}$ during the same epoch, indicating that the binary becomes significantly less gravitationally bound (Figure \ref{Figure9b}).

The simultaneous decrease in orbital angular momentum and increase in orbital energy reflects two competing effects. The loss of mass from the system removes angular momentum, leading to a decline in J$_{orb}$. However, because a substantial fraction of the mass is lost from the binary rather than accreted, the total mass of the system decreases significantly, reducing its gravitational binding. This causes the orbital energy to increase (become less negative), indicating that the system becomes progressively less bound. As a result, the orbital separation increases, leading to orbital expansion despite the ongoing loss of angular momentum. This behavior is a characteristic signature of highly non-conservative mass transfer. Following the formation of the WD, the orbital energy remains nearly constant, and at the present epoch ($t = 1.720\,\mathrm{Gyr}$), the system maintains $E_{\rm orb} = -2.0\times10^{45}\,\mathrm{erg}$, confirming that the binary has settled into a stable, detached BSS+WD configuration.

\begin{figure*}[h]
\centering
\includegraphics[width=0.6\textwidth]{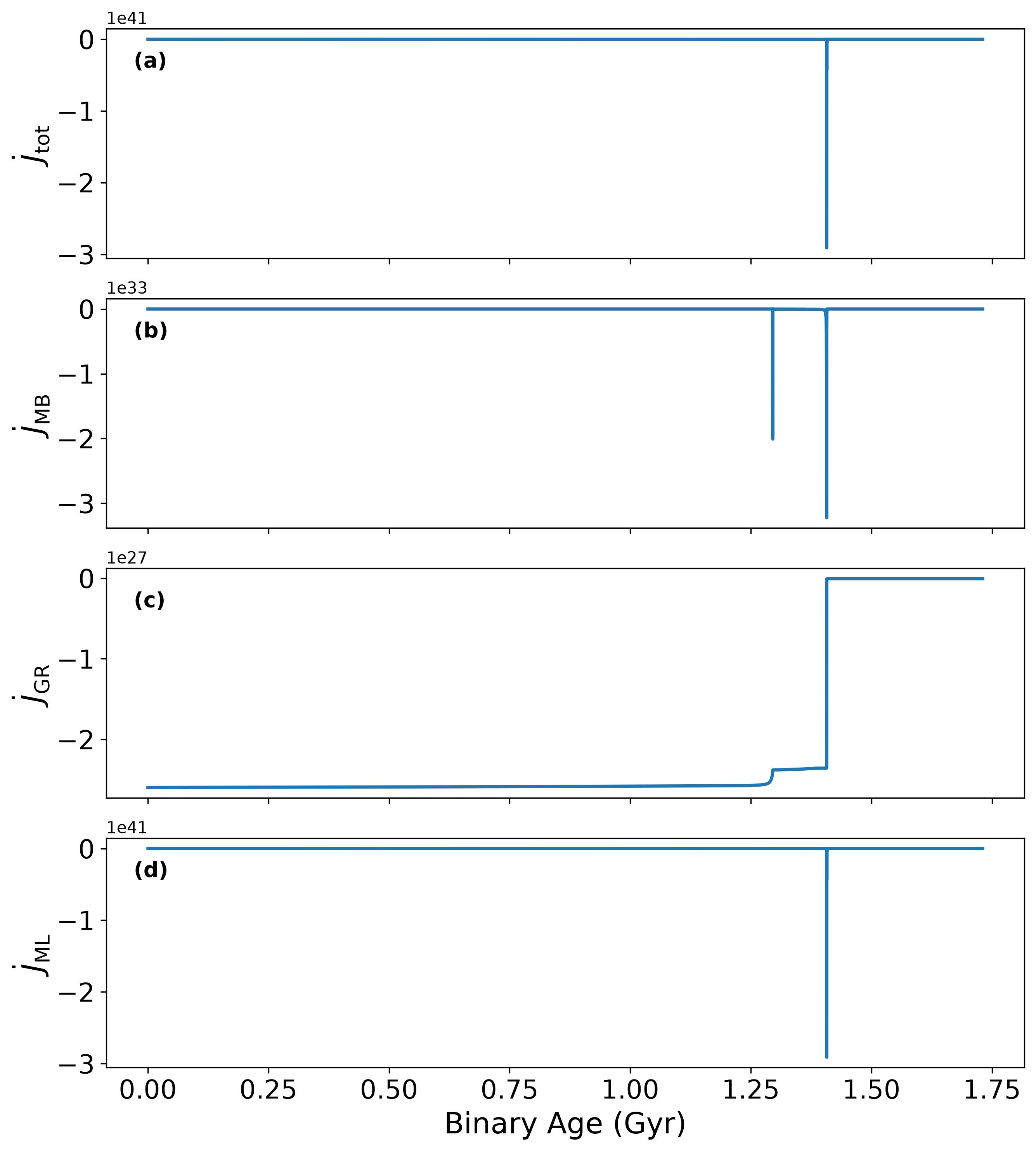}
\caption{Evolution of the angular momentum loss rates as a function of binary age. Panels show the total loss rate and the contributions from magnetic braking, gravitational radiation, and systemic mass loss. The angular momentum evolution is dominated by systemic mass loss during the TP-AGB phase, while magnetic braking plays a minor role and gravitational radiation remains negligible throughout. Note that each panel uses an independent vertical
scale, with the corresponding multiplicative factor indicated on the
vertical axis.}
\label{Figure10}
\end{figure*}

To identify the dominant mechanisms responsible for the evolution of the orbital angular momentum, we examine the individual contributions from different angular momentum loss mechanisms, as shown in Figure \ref{Figure10}. The total angular momentum loss rate, $\dot{J}_{\rm tot}$, is dominated by the systemic mass-loss term, $\dot{J}_{\rm ML}$, throughout the binary evolution. Although magnetic braking and gravitational radiation are included in the calculations, their contributions remain several orders of magnitude smaller than the dominant systemic mass-loss contribution ($|\dot{J}_{\rm ML}|\sim3\times10^{41}$, $|\dot{J}_{\rm MB}|\sim10^{33}$, and
$|\dot{J}_{\rm GR}|\sim10^{27}\,\mathrm{g\,cm^{2}\,s^{-2}}$)
and therefore have a negligible effect on the orbital evolution.
Consequently, the angular momentum evolution of WOCS~5011 is governed almost entirely by mass loss from the system. A significant transition occurs during the TP-AGB phase at $t = 1.39 \,\mathrm{Gyr}$, where the angular momentum loss rates exhibit sharp negative spikes reaching $|\dot{J}_{\rm tot}| \sim (1$--$3)\times10^{41}\,\mathrm{g\,cm^{2}\,s^{-2}}$. These pronounced dips correspond to brief but intense episodes of mass loss from the donor star, during which a large fraction of the stellar envelope is expelled from the system. The escaping material carries away a substantial amount of orbital angular momentum, producing the dominant contribution seen in $\dot{J}_{\rm ML}$. This phase represents the key evolutionary stage in which the binary undergoes rapid angular momentum removal, directly driving the sharp decline in the orbital angular momentum.

\subsection{Revisiting the spectral energy distribution of WOCS 5011}

\citet{vaidya2022uocs} performed SED analysis of the BSSs in NGC~7789 observed with UVIT, including WOCS~5011. The SED of WOCS~5011 was satisfactorily fitted with a single stellar component with $T_{\rm eff}=8750$~K. The details of the photometric data is described in \citet{vaidya2022uocs}.

We revisited the SED of WOCS~5011 in view of the WD companion predicted by the MESA binary-evolution model. For a lower BSS temperature of $T_{\rm eff}=8250$~K, the expected FUV contribution from the BSS is lower, making the excess in the UVIT/F169M band more apparent. We therefore performed a two-component SED fit using the Python-based BINARY SED FITTING code of \citet{jadhav2021blue}. A stellar atmosphere model was used for the BSS component, while Koester WD atmosphere models \citep{koester2002white} were used for the hot component. The best-fitting model was obtained through $\chi^{2}$ minimization, following the SED-fitting procedure described in \citet{vaidya2022uocs, panthi2022uocs, panthi2026uocs}.

\begin{figure*}[h]
\centering
\includegraphics[width=0.5\linewidth]{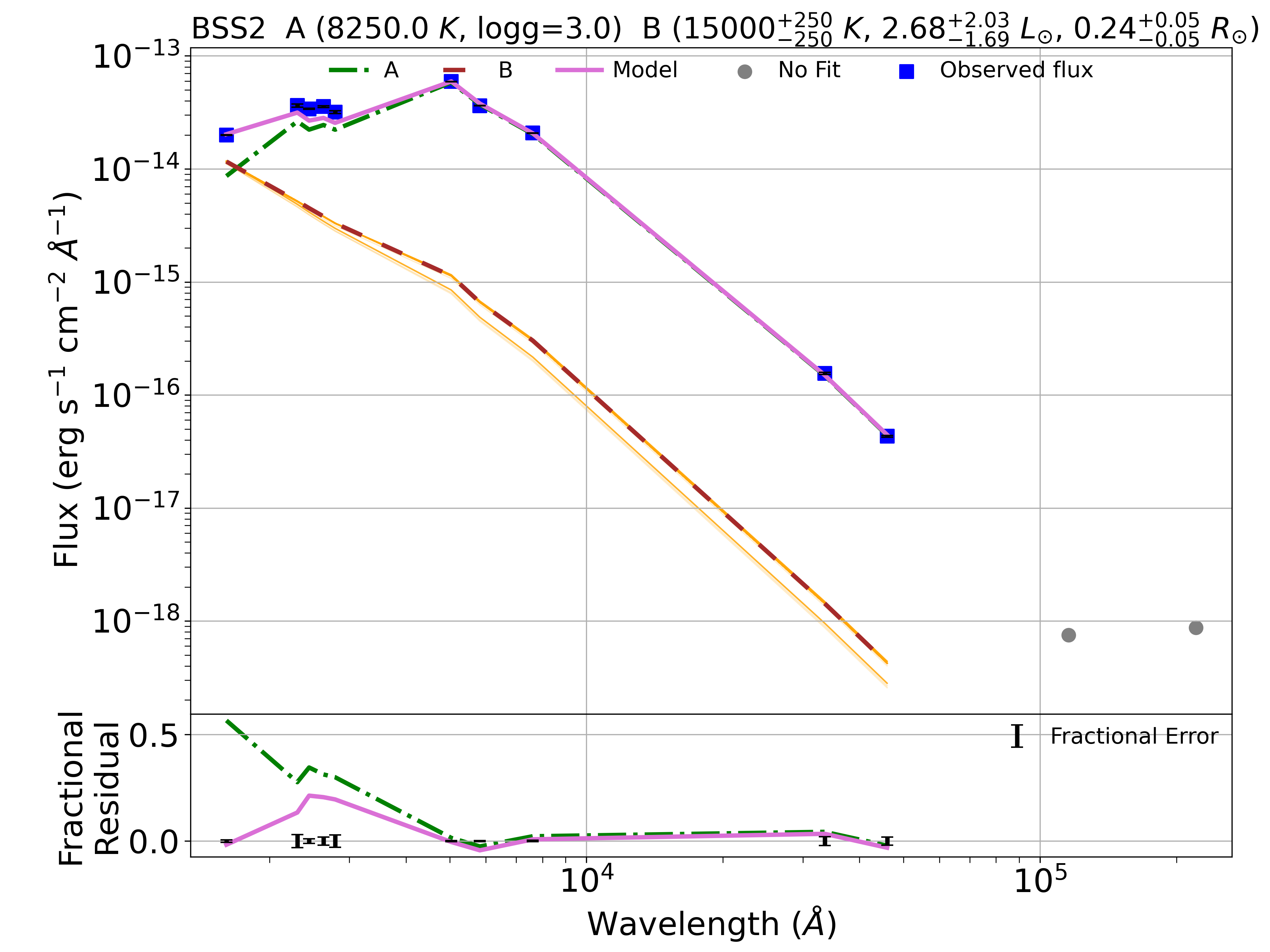}
\caption{Two-component spectral energy distribution of WOCS~5011. The observed fluxes are shown as blue squares, while the gray circles represent photometric points excluded from the fit. The green dash-dotted curve represents the BSS component (A), the brown dashed curve represents the hot component (B) fitted using a Koester WD atmosphere model, and the 
magenta curve represents the combined model. The lower panel shows the fractional residuals between the observed and model fluxes.}
\label{Figure11}
\end{figure*}

The resulting two-component SED is shown in Figure~\ref{Figure11}. The fit yields a hot component with $T_{\rm eff}=13,250\pm250$~K, $L=2.86^{+2.19}_{-1.81}\,L_{\odot}$, and 
$R=0.32\pm0.07\,R_{\odot}$. The temperature obtained from the SED fit is comparable to the $T_{\rm eff}=12,520$~K predicted for the WD by the MESA model. However, the SED-derived radius and luminosity are substantially larger than the MESA predictions of 
$R=0.012\,R_{\odot}$ and $L=0.003\,L_{\odot}$.

\section{Discussion} \label{Section5} \label{Section5}

Binary population-synthesis studies have long struggled to reproduce the observed numbers and orbital properties of BSSs in old OCs \citep{tian2006blue, lu2011modeling, geller2013direct}. These models also tend to produce too many high-mass BSSs and too few objects near the turnoff, suggesting that the treatment of mass-transfer efficiency and stability remains incomplete \citep{leiner2021census}. The evolutionary modeling presented in this work provides a physically consistent formation pathway for the BSS WOCS~5011 through binary interaction. In the best-fit model the initially more massive star evolves through the RGB and AGB phases, transfers material to the initially less massive companion, and ultimately becomes a WD, while the accretor forms a BSS. The resulting present-day system is a BSS+WD binary with an orbital period of $\sim2708$ days from the best-fit model, in excellent agreement with the observed value of $2710\pm60$ days from RV monitoring.

The chemical properties of WOCS~5011 provide an additional and independent line of evidence for a post-AGB mass-transfer origin. The Ba enhancement reported in literature for WOCS~5011 is naturally interpreted as the signature of past accretion from an evolved AGB donor. Because the present BSS cannot synthesize such heavy elements internally at its current evolutionary stage, the observed Ba overabundance must have been acquired externally. 
Binary evolution calculations have shown that stable RLOF from RGB donors can produce post-mass-transfer systems with orbital periods extending to several hundred days and, under favorable conditions, approaching $\sim10^{3}$ days \citep{woods2011can,ge2015adiabatic,ge2020adiabatic}. At wider separations, however, stable RLOF becomes increasingly difficult to sustain, and such systems are more naturally explained by wind-driven mass transfer from AGB donors \citep{izzard2009population,abate2013wind,abate2015modelling}. At wider separations, systems with periods exceeding $\sim4000$ days have been interpreted as products of AGB WRLOF \citep{abate2015modelling}. The observed orbital period of 2710 $\pm$ 60 days in WOCS~5011 is reproduced in our model through a wind-driven mass-transfer history in which WRLOF operates over an extended phase from the RGB through the AGB, while a short-lived but dominant RLOF episode during the TP--AGB contributes the bulk of the accreted mass. 

The eccentricity--period diagram of Ba stars compiled by \citet{jorissen2019barium} shows that many such systems occupy the intermediate- to long-period regime ($10^{3}$--$10^{4}$ days) with moderate eccentricities. With an orbital period of 2710 days and an eccentricity of 0.32, WOCS~5011 falls naturally within this region, consistent with binaries that have experienced mass transfer from an evolved AGB donor while avoiding strong tidal circularization. Likewise, in the orbital period--WD mass diagram presented by \citet{mathieu2025blue}, the location of WOCS~5011 at $P_{\rm orb}\sim2700$~days and $M_{\rm WD}\sim0.55\,M_{\odot}$ places it among systems expected to have formed from comparatively evolved donors, again supporting a TP--AGB mass-transfer origin.

We note that the present calculations assume a circular orbit, whereas WOCS 5011 has a present-day eccentricity of \(e=0.32\pm0.04\). As discussed in Section 3.1, eccentricity can affect the orbital separation and Roche-lobe geometry, potentially altering the mass-transfer history and the subsequent evolution of the donor remnant. Therefore, the circular-orbit approximation may contribute to the differences between the SED-inferred radius and luminosity of the hot companion and those predicted for the final WD by the MESA model. Quantifying this effect would require binary evolution calculations that explicitly account for eccentricity. The evolution of stellar spins is also not included in the present calculations.

The successful reproduction of the observed properties of WOCS~5011, supports the view that mass transfer from TP--AGB donors is an important formation pathway for BSSs in OCs. More broadly, this system highlights the importance of interaction regimes that combine wind accretion, WRLOF, and brief direct overflow, and therefore provides an important benchmark for future binary evolution and population-synthesis studies.

\section{Summary and Conclusions} \label{Section6}

In this work, we have presented a comprehensive binary evolution model for the formation of the binary BSS WOCS 5011 in the OC NGC~7789. Using detailed \texttt{MESA} simulations, we followed the evolution of both stars in the progenitor binary from the zero-age main sequence to the present-day detached configuration, and successfully reproduced the observed physical and orbital properties of the system. The progenitor begins as a wide binary consisting of a $1.85\,M_{\odot}$ primary star and a $1.67\,M_{\odot}$ companion.

In this study, the evolutionary pathway detailed by the model to match the observations throws light on various details that are crucial to understand the binary evolution as well as the processes that shape the outcome. The mass-transfer history of the system reveals that multiple channels contribute to the growth of the BSS. Wind mass transfer during the RGB phase initiates the process, while WRLOF during the AGB phase provides a dominant contribution to the total accreted mass. Regular RLOF occurs only during the largest thermal pulses but contributes significantly due to its high efficiency. This combination of processes includes non-conservative mass transfer, in which a substantial fraction of the donor’s mass is lost from the system. The orbit expands throughout most of the evolution, particularly during the AGB phase, due to systemic mass transfer and mass ratio reversal. This expansion plays a crucial role in maintaining the stability of the mass-transfer process \citep{sun2023wocs}. The mass-transfer phase is relatively short compared to the total evolutionary timescale, indicating that the formation of the BSS is a recent event in the history of the system.

The modeled BSS has $T_{\rm eff}=8409$~K, $R=3.27\,R_{\odot}$, and $L=47.8\,L_{\odot}$, consistent with the observational estimates of $8350\pm140$~K \citep{nine2024wiyn}, a radius of $3.04\pm0.15\,R_{\odot}$ \citep{vaidya2022uocs}, and a luminosity of $47.7\pm4.8\,L_{\odot}$ \citep{vaidya2022uocs}.
The surface gravity predicted by the model, $\log g = 3.75$, is also consistent with the observed value of $3.66\pm0.07$ \citep{nine2024wiyn}. The final mass of the accretor, $\sim2.17\,M_{\odot}$, is well above the cluster turnoff mass, showing that substantial accretion and rejuvenation have occurred. The model further predicts that the former donor has evolved into a CO WD with a mass of $\sim0.55\,M_{\odot}$, an effective temperature of $\sim12500$~K, a radius of $\sim0.0136\,R_{\odot}$, and a luminosity of only $\sim0.004\,L_{\odot}$.
Motivated by the prediction of WD from MESA, we revisited the SED of WOCS~5011. The two-component SED fit gives a BSS with $T_{\rm eff}=8250$~K, $R=3.24\pm0.16\,R_{\odot}$, and $L=43.77\pm4.40\,L_{\odot}$, in good agreement with the observed and modeled properties of the BSS. The hot component has $T_{\rm eff}=13250\pm250$~K, $R=0.32\pm0.07\,R_{\odot}$, and 
$L=2.86^{+2.19}_{-1.81}\,L_{\odot}$. While its temperature is close to the MESA prediction, the SED-derived radius and luminosity are substantially larger than those predicted by MESA for the compact WD. This discrepancy may indicate limitations in the WD models when applied to recently stripped systems. We note, however, that the present modeling assumes a circular orbit and does not include the evolution of orbital eccentricity or stellar spins. In particular, the circular-orbit approximation could contribute to the differences between the inferred properties of the hot companion and those predicted by the MESA model.
These quantities may provide further constraints in future work.

\section*{Acknowledgments}

The Authors would like to thank the referee for the valuable comments, that has improved the manuscript.
We thank the MESA developers and community for making the code publicly
available and for their continued development and support of the project.
This research made use of Modules for Experiments in Stellar Astrophysics
(MESA; \citealt{paxton2011modules,paxton2013modules,paxton2015modules,paxton2019modules}), as well as the Python packages
NumPy \citep{harris2020array}, SciPy \citep{virtanen2020scipy},
pandas \citep{mckinney2010data}, and Matplotlib
\citep{hunter2007matplotlib}.

\section*{Data Availability}

The MESA inlists and associated Fortran codes used in this work are publicly
available on Zenodo at \url{https://zenodo.org/records/21923458}. The
observational constraints used in the binary evolution modeling are taken
from the published literature and are referenced throughout the manuscript.

%\subsection{Tables} \label{subsec:tables}
\bibliography{sample701}{}
\bibliographystyle{aasjournalv7}

\end{document}